\documentclass[twocolumn,showpacs,nofootinbib,preprintnumbers,amssymb]{revtex4-1}

\def\pmb#1{\setbox0=\hbox{$#1$}%
  \kern-.025em\copy0\kern-\wd0
  \kern.05em\copy0\kern-\wd0
  \kern-.025em\raise.0433em\box0}

\usepackage{latexsym}
\usepackage{gensymb, graphicx,epsf, epsfig, amssymb}% Include figure files
\usepackage{bm}
\usepackage{booktabs,ctable}

\def\be{\begin{equation}}
\def\ee{\end{equation}}
\def\beq{\begin{eqnarray}}
\def\eeq{\end{eqnarray}}

\newcommand{\pxx}{\partial^{2}_{xx}}
\newcommand{\pyy}{\partial^{2}_{yy}}
\newcommand{\pzz}{\partial^{2}_{zz}}
\newcommand{\pxy}{\partial^{2}_{xy}}

\newcommand{\ptt}{\partial^{2}_{tt}}

\newcommand{\pxt}{\partial^{2}_{xt}}
\newcommand{\pyt}{\partial^{2}_{yt}}

\newcommand{\px}{\partial_{x}}
\newcommand{\py}{\partial_{y}}

\newcommand{\vo}{V^{(0)}}
\newcommand{\voo}{V^{(2)}}
\newcommand{\vx}{V^{(0)}_{x}}
\newcommand{\vy}{V^{(0)}_{y}}

\newcommand{\vxx}{V^{(1)}_{x}}
\newcommand{\vyy}{V^{(1)}_{y}}

\begin{document}

\title{Tidal interaction in compact binaries: a post-Newtonian affine framework}

\author{V.~Ferrari, L.~Gualtieri, A.~Maselli}
\address{Dipartimento di Fisica, ``Sapienza'' Universit\`a di Roma
  \& Sezione INFN Roma1, Piazzale Aldo  Moro 5, 00185, Roma, Italy}

\begin{abstract} 
  We develop a semi-analytical approach, based on the post-Newtonian expansion
  and on the affine approximation, to model the tidal deformation of neutron
  stars in the coalescence of black hole-neutron star or neutron star-neutron
  star binaries. Our equations describe, in a unified framework, both the system
  orbital evolution, and the neutron star deformations. These are driven by the
  tidal tensor, which we expand at $1/c^{3}$ post-Newtonian order, including
  spin terms.  We test the theoretical framework by simulating black
  hole-neutron star coalescence up to the onset of mass shedding, which we
  determine by comparing the shape of the star with the Roche lobe.  We validate
  our approach by comparing our results with those of fully relativistic,
  numerical simulations.
\end{abstract}

\maketitle

%%%%%%%%%%%%%%%%%%%%%%%%%%%%%%%%%%%%%%%%%%%%%%%%%%%%%%%%%%%%%%%%%%%%%%%%%%%%%%%%%%%%%%%%
\section{Introduction}\label{intro}
%%%%%%%%%%%%%%%%%%%%%%%%%%%%%%%%%%%%%%%%%%%%%%%%%%%%%%%%%%%%%%%%%%%%%%%%%%%%%%%%%%%%%%%%
Coalescing binaries composed of neutron stars (NS) and/or black holes (BH), are
among the most promising sources of gravitational waves (GWs) to be detected by
gravitational wave interferometers like Virgo and LIGO \cite{virgoligo}. These
systems are also interesting since they are thought to be related to short
gamma-ray burst \cite{NPP09}.

The process of coalescence has been studied mainly using post-Newtonian (PN) and
fully relativistic, numerical simulations.  PN expansion \cite{B06} has the
advantage of providing a semi-analytic description of the evolution of the
system, but it is poorly convergent in the strong field limit; therefore it is
appropriate to study the inspiral phase only. These limitations have been
overcome by PN resummed formulations like EOB \cite{EOB}, but these
formulations are not able to describe the dynamical features of the stellar
deformation.
Moreover, in the standard PN expansion the compact objects are
treated as pointlike up to the 4.5 (included) post-Newtonian order.  Finite size
effects are, formally, of order 5PN; however their contribution is larger than
what a naive counting of PN orders may suggest \cite{MW04}.  Tidal deformations
have recently been included in the PN framework through the ``Love number''
approach \cite{FH08,H08}, which assumes that the tidal field is proportional to
the quadrupole momentum (see below). The effects of the tidal deformation
on the orbital motion have been studied in \cite{VFH11,EOBtidal}.

Fully relativistic codes are the most powerful tool to investigate the latest
phases of the inspiral and merger (see \cite{F09} for a review on the
subject). They are, however, not exempt from drawbacks: their computational cost
is high, therefore the parameter space cannot be explored at large; furthermore,
initial data solvers are still unable to provide accurate initial data for
binaries with non-aligned spins, and may introduce spurious numerical effects
which, if not appropriately cured, affect the subsequent evolution of the
system.  These problems are of particular relevance in BH-NS binaries, where the
lack of symmetry makes more difficult to follow the entire process of
coalescence by fully relativistic simulations.  For these reasons, the process
has been studied in the literature using some simplifying assumptions, or for a
restricted set of parameters. For instance, in \cite{TBFS07,S09q,FGP09,FGP10}
the inspiral is modeled as a sequence of quasi-equilibrium circular orbits with
decreasing radius; in \cite{TBFS07,S09q} the process is studied by fully
relativistic simulations, whereas \cite{FGP09,FGP10} use the affine approach
(see below).  In \cite{F06a,F06b} Einstein's equations are evolved assuming that
the black hole is non rotating, and for large values of the mass ratio
$q=M_{BH}/M_{NS}$, whereas in \cite{S07,S09,S10,E09} $q$ takes values $q\le5$;
in \cite{E09,S11} the black hole is assumed to rotate with spin parallel to the
orbital angular momentum, and different values have been considered. For a
recent review on fully relativistic simulations of BH-NS binaries see
\cite{ST11} (the literature on NS-NS coalescing binaries is much more extended,
and we do not report it here).

In this paper we develop a semi-analytic approach to study BH-NS and NS-NS
coalescence, by merging two different frameworks: the PN approach, which
accurately describes the system orbital motion, and the affine model
\cite{CL85,LM85,WL00,CFS06,FGP09,FGP10}, which describes the stellar
deformations induced by the tidal field.  To this aim, we compute the tidal
tensor associated to the PN metric of a two-body system, defined in terms of the
PN Riemann tensor and of the local tetrad of the deformed body
\begin{equation}
C_{(i)(j)}=R_{\alpha\beta\gamma\delta}e^{\alpha}_{(0)}e^{\beta}_{(i)}e^{\gamma}_{(0)}e^{\delta}_{(j)}\,,
\label{defC}
\end{equation}
up to $O(1/c^3)$. This tensor was derived with a different approach in
\cite{DSX92,RF05,VF09} up to $O(1/c^2)$; our expression coincides with that of
\cite{VF09} and also includes $O(1/c^3)$ terms, associated to the spins of the
compact objects.

In the affine model the NS is described as a deformable ellipsoid, subject to
its self-gravity, to internal pressure forces and to the tidal field of the
companion. In the original formulation of this approach, the NS structure was
considered at a Newtonian order, assuming a polytropic equation of state (EOS)
\cite{CL85,LM85,WL00,CFS06}. A first improvement was introduced in
\cite{FGP09,FGP10}, where general relativity was taken into account in the
description of the stellar structure, and non-polytropic EOSs were
considered. This approach was used to study quasi-equilibrium configuration
sequences of BH-NS systems, in order to estimate the critical distance at which
the NS is disrupted by the tidal interaction \cite{FGP09}, to determine the
corresponding cut-off frequency in the emitted gravitational wave signal
\cite{FGP10,PROR11}, and to estimate the mass of the torus which forms after the
NS is disrupted \cite{PTR11}.

In this paper we further improve the affine model introducing a more accurate
description of the orbital motion and of the tidal interaction.  Our approach
differs from existing work on NS tidal deformation in compact binaries in the
following aspects.
\begin{itemize}
\item In \cite{WL00,CFS06,FGP09,FGP10,OK96,PTR11} the affine model was used
  assuming that the NS follows a timelike geodesic of Kerr's spacetime; this
  approximation fails when the mass ratio $q$ is low. In addition, time-dilation
  factors were neglected.

  Furthermore, most works employing the affine model
  \cite{WL00,CFS06,FGP09,FGP10} do not evolve the dynamical equations of stellar
  deformation. Rather, they find, at each value of the orbital radius, the
  corresponding stationary configuration describing the deformed star. In
  \cite{OK96,PTR11} the dynamical equations were solved; however, while the
  orbital evolution was described in PN coordinates, the BH tidal field was
  expressed in the Boyer-Lindquist coordinates of the Kerr metric describing a
  single BH; neglecting the difference between these coordinate systems yields a
  loss of accuracy of the model.

  These problems are solved in the fully consistent approach presented in this
  paper, where the BH-NS or NS-NS systems are described by a two-body PN metric,
  {\it which holds for any value of the mass-ratio $q$}. The tidal tensor itself
  is expressed in the PN coordinates, and the proper time of each compact body
  is expressed in terms of the PN time coordinate, through the appropriate
  Lorentz factor. Our approach is valid up to the onset of mass shedding, which
  occurs when the deformed star crosses the Roche lobe; after that, it can no
  longer be applied, since the assumption that the star is a deformed ellipsoid
  is significantly violated.  We describe the orbital motion of the compact
  objects using the $3.5$ PN equations for pointlike objects, with
  next-to-leading-order tidal corrections; the NS tidal deformation is driven by
  the tidal tensor of the $3$ PN metric.  The dynamical equations are a system
  of (non-linear) ordinary differential equations in time.

\item NS tidal deformations have also been studied in a series of paper
  \cite{FH08,H08,HLLR10,DN09,BP09,DN10}, where the deformation properties have
  been encoded in a set of numbers, the {\it Love numbers}, which relate the
  quadrupole tensor (or, more generally, the multipole moments of the star) to
  the tidal tensor.  This approach is grounded on the {\it adiabatic
    approximation}, i.e., on the assumption that the orbital evolution timescale
  is much larger than the timescale needed for the star to set into a stationary
  configuration. In this approximation, the quadrupole tensor is proportional to
  the tidal tensor:
\begin{equation}
  Q_{(i)(j)}=\lambda C_{(i)(j)}\,,\label{ad}
\end{equation}
with $\lambda$ constant.  The Love number $\lambda$ can be computed by studying
the response of a single star to an external tidal tensor
\cite{FH08,H08,DN09,BP09}.  This model has been employed to determine the effect
of tidal deformation on the orbital motion of a NS in a binary system
\cite{HLLR10,VF09,DN10,VFH11}.

We also compute the Love number $\lambda$ (see Section~\ref{validate}), without
assuming the adiabatic approximation: the stellar deformation is found by
solving dynamical equations.
\end{itemize}
To test the accuracy of our approach, we compare the results with the existing
literature on BH-NS binaries.  As a preliminary check, we verify that our PN
description of the orbital motion accurately reproduces the fully relativistic
results \cite{S09,E09,privcomm}.  Then, we verify that the onset of mass
shedding we determine, is consistent with the results of fully relativistic
simulations \cite{S09,E09,privcomm}.  Finally, we check that the stellar
deformations predicted by our model are consistent with existing computations of
the Love number \cite{FH08,H08,DN09,BP09}.

This paper focuses mainly on the theoretical framework, and on its validation by
comparison with the existing literature, where available.  The tool we develop
will be used in future works to study the dynamics of compact binaries.

The plan of the paper is the following. In Sec~\ref{model} we describe
the model. In Sec.~\ref{validate} we assess the validity of our approach, by
comparing the results with the literature. In Sec.~\ref{conclusions} we draw the
conclusions.
%%%%%%%%%%%%%%%%%%%%%%%%%%%%%%%%%%%%%%%%%%%%%%%%%%%%%%%%%%%%%%%%%%%%
\section{The model}\label{model}
%%%%%%%%%%%%%%%%%%%%%%%%%%%%%%%%%%%%%%%%%%%%%%%%%%%%%%%%%%%%%%%%%%%%
We use notations and conventions introduced in \cite{FGP09}, where the affine
model (partially improved with respect to the original formulation
\cite{CL85,LM85}) is widely discussed. In the following $m_{1},m_{2}$ are the
masses of the two compact objects; we shall consider the tidal deformation of
the NS with mass $m_1$ and radius $R_{NS}$; the companion, with mass $m_2$, can
either be a BH or a NS. Furthermore, we define $m=m_{1}+m_{2}$ and
$\nu=\mu/m=m_{1}m_{2}/m^{2}$, and the mass ratio $q=m_2/m_1$.

%%%%%%%%%%%%%%%%%%%%%%%%%%%%%%%%%%%%%%%%%%%%%%%%%%%%%%%%%%%%%%%%%%%%%
\subsection{Improved  affine model} \label{sec:affine}
%%%%%%%%%%%%%%%%%%%%%%%%%%%%%%%%%%%%%%%%%%%%%%%%%%%%%%%%%%%%%%%%%%%%%
The basic assumption of the affine model is that the NS, deformed by the tidal
field, maintain an ellipsoidal shape; it is an $S$-type Riemann ellipsoid, i.e.,
its spin and vorticity are parallel, and their ratio is constant \cite{EFE}.
The deformation equations are written in the {\it star principal frame}, i.e.,
the frame comoving with the star, whose axes coincide with the ellipsoid
principal axes.  In what follows, $a_1$ is the axis which points toward the
companion; $a_2$ and $a_3$ are the axes orthogonal to $a_1$, with $a_2$ lying in
the orbital plane; the indices 1,2,3 label the corresponding directions.
Surfaces of constant density inside the star form self-similar ellipsoids and
the velocity of a fluid element is a linear function of the coordinates $x^{i}$
in the principal frame. Under these assumptions, the infinite degrees of freedom
of the stellar fluid motion can be reduced to five \cite{CL85,LM85,WL00}, and
are associated to dynamical variables governed by a set of non-linear
differential equations, which describe the evolution of the stellar
deformation. These variables are the three principal axes of the ellipsoid
$a_{i}$ $(i=1,2,3)$ and two angles, $\psi$, $\lambda$, defined as
\begin{equation}\label{NSspin}
\frac{d\psi}{d\tau}=\Omega\ ,\qquad \frac{d\lambda}{d\tau}=\Lambda\ , 
\end{equation}
where $\tau$ is the NS proper time, and $\Omega$ is the ellipsoid
angular velocity,
measured in the parallel transported frame associated with the star center of
mass. $\psi$ is the angle between the principal frame and the parallel
transported frame.   $\Lambda$ is defined as follows:
\begin{equation}
\Lambda=\frac{a_{1}a_{2}}{a_{1}^{2}+a_{2}^{2}}\zeta\ ,
\end{equation}
where $\zeta$ is the vorticity along the axis $x^3$ in the principal frame.
The NS internal dynamics is described in terms of the Lagrangian
\begin{equation}\label{lagrangeI}
\mathcal{L}_{I}=T_{I}-U-{\cal V}~,
\end{equation}
where $T_{I}$ is the fluid kinetic energy, $U$ is the internal energy and ${\cal
  V}$ is the star self-gravity.  In the original approach introduced by Carter
and Luminet, these are defined in a Newtonian framework
\begin{eqnarray}
T_I&=&\frac{1}{2}\int v^{2} dM_{B}~,\label{lag0}\\
U&=&\int\frac{\epsilon}{\rho}dM_{B}~,\label{lag1}\\
{\cal V}&=&-\frac{G}{2}\int\frac{dM_{B}dM_{B}'}{\vert{\bf x}-{\bf x}'\vert}
=\int dM_B r\partial_{r}\Phi_{Newt}\,,
\label{lag}
\end{eqnarray}
where $M_B$ is the baryonic mass, $\rho$ the mass density, $\epsilon$ the
Newtonian energy density, $\Phi_{Newt}$ the gravitational potential; all these
quantities are solutions of the Newtonian equations of stellar structure.
Furthermore, $dM_B=\rho d^3x$ and ${\cal V}$ satisfies the virial theorem, which
states that, in the spherical configuration, ${\cal V}=-3\Pi$, where
\begin{equation}
\Pi=\int \frac{p}{\rho}dM_B=\int pd^3x\,.\label{dfpi}
\end{equation}
The variation of the Lagrangian (\ref{lagrangeI}) gives the equations of motion
for the five dynamical variables $a_i,\,\psi,\,\lambda$.

In \cite{CL85,LM85} it was shown that, under the affine hypothesis, the
integrals in Eqns.~(\ref{lag0}-\ref{lag}) and their variations, can be expressed
in terms of integrals on the fluid variables computed for the spherical
configuration ($a_i=R_{NS}$), and of functions of the dynamical variables.  With
this simplification, the equations of motion can easily be found.  In the
following, a superscript hat will denote quantities computed for the spherical
star.  $T$ and ${\cal V}$ expressed in terms of the ``hatted'' quantities and of
the dynamical variables are
\begin{eqnarray}
&&T_{I}=\sum_{i}\frac{1}{2}\left(\frac{da_i}{d\tau}\right)^{2}
 \frac{\hat{\cal M}}{R_{NS}^2}\nonumber\\
&&+\frac{1}{2}\frac{\hat{\cal M}}{R_{NS}^2}\left[\left(\frac{a_{1}}{a_{2}}\Lambda-\Omega\right)^{2}\!\!a_2^2\
\!+\!\left(\Omega-\frac{a_{2}}{a_{1}}\Lambda\right)^{2}\!\!a_1^2\right]\\
&&{\cal V}=\frac{1}{2}\hat{{\cal
V}}R_{NS}\int^{\infty}_{0}\frac{d\sigma}{\sqrt{(a_{1}^{2}+\sigma)(a_{2}^{2}+\sigma)(a_{3}^{2}+\sigma)}}
,\label{defV}
\end{eqnarray}
where  $\hat{\cal M}$ is the {\it scalar quadrupole moment}
\begin{equation}
\hat{\cal M}=\frac{1}{3}\int_{sph} \sum_i (x_i)^2 dM_B \label{defmhat}
\end{equation}
(the subscript $sph$ means that the integration is performed on the spherical
star) and $\hat{\cal V}$ is the self-gravity potential of the spherical star.

The procedure to make explicit the dependence of the internal energy $U$ on the
dynamical variables is more subtle.  The internal energy variation $dU$ can be
written as
\begin{equation}
dU=\sum_{i}\frac{\Pi}{a_{i}}da_{i}\,.
\label{defU}
\end{equation}
The pressure integral $\Pi$ given by Eq.~(\ref{dfpi}) can not be factorized in a
spherical integral and a function of the axes; however, it can be expressed as
\begin{equation}
\Pi=\frac{a_1a_2a_3}{R_{NS}^3}\int_{sph}p(\rho)d^3 x~,
\end{equation}
where $p(\rho)$ is the fluid equation of state, and $\rho$ is the
rescaled mass density 
\begin{equation}
\rho=\hat\rho\frac{R_{NS}^3}{a_1a_2a_3}
\end{equation}
with $\hat\rho$ mass density in the spherical configuration. When $a_i=R_{NS}$,
the pressure integral $\Pi$ reduces to the spherical pressure integral
$\hat\Pi$.

A first improvement to this approach was introduced in \cite{FGP09}, where the
Newtonian description of the NS equilibrium configuration was replaced by the
relativistic equations of stellar structure (TOV)
\begin{eqnarray}
\partial_{r_s}m_s&=&4\pi\hat\epsilon r_s^2\nonumber\\
\partial_{r_s}\hat p&=&-G\frac{(\hat\epsilon+\hat p/c^2)(m_s+4\pi
\hat pr_s^3/c^2)}{r_s(r_s-2Gm_s/c^2)}\label{toveq}~.
\end{eqnarray}
Here $\hat\epsilon$ is the relativistic mass-energy density in the spherical
configuration, $r_s$ is the radial coordinate in a Schwarzschild frame
associated to the non rotating NS, and $m_s(r_s)$ is the gravitational mass
enclosed in a sphere of radius $r_s$. We remark that this is a major change,
since the relativistic radius is smaller than the Newtonian radius by
$\sim10\%-20\%$. We also remark that the Schwarzschild coordinate $r_s$ is
different from the radial coordinate in the Newtonian frame
$r=\sum_i(x_i^2)^{1/2}$.  The self-gravity integral $\hat V$ was changed
accordingly, as
\begin{equation}
\hat{\cal V} =\int_{sph} dM_B r_s\partial_{r_s}\Phi_{TOV}\,,
\end{equation}
where $\Phi_{TOV}$ is an effective relativistic gravitational potential of the
spherical star, defined in terms of the TOV equations as follows
\begin{equation}
\hat\rho\partial_{r_s}\Phi_{TOV}=G\frac{(\hat\epsilon+\hat p/c^2)(m_s+4\pi
r_s^3\hat p/c^2)}{r_s(r_s-2Gm_s/c^2)}\,.\label{defphitov}
\end{equation}
With this definition the virial theorem
\begin{equation}
\int_{sph} r_s\partial_{r_s}\Phi_{TOV}dM_B =-3\int_{sph} \frac{\hat p}{\hat\rho}dM_B\label{virial}
\end{equation}
is still satisfied, with $\hat p$ solution of the TOV equations (\ref{toveq})
and $\hat\rho$ baryon mass density.  As shown in Section \ref{sec:intdyn}, some
terms in the dynamical equations cancel in the spherical limit, only if the
virial theorem is satisfied. A non-exact cancellation of these terms would lead
to strong instabilities.

A further improvement, which we introduce in this paper, consists in a careful
treatment of the coordinate frames. To describe the integrals in the spherical
configuration, the relevant coordinate systems are: (i) the Schwarzschild frame,
with radial coordinate $r_s$, in which the TOV equations (\ref{toveq}) are
expressed; (ii) the Newtonian frame for a spherical star $\{x^i\}$, which we now
replace with the corresponding $1$ PN post-Newtonian frame \cite{B06}, with
isotropic radial coordinate $r={\sum_i(x_i^2)}^{1/2}$ and metric (for a single
star)
\begin{equation}
ds^2=-\left(1-\frac{2V}{c^2}+\frac{2V^2}{c^4}\right)dt^2+
\left(1+\frac{2V}{c^2}\right)\delta_{ij}dx^idx^j\,,\label{1PNstar}
\end{equation}
where $V(r)\equiv G\int^{\infty}_{r}\frac{m_s(r')}{r^{'2}}d{r}'$.  Following
\cite{Chan65} the transformation between the post-Newtonian isotropic radial
coordinate and the Schwarzschild coordinate inside the star is given by
\begin{equation}
r=r_s\left(1-\frac{V(r_s)}{c^2}\right)\label{rsr}\,.
\end{equation}
The scalar quantity $dM_B$ can be expressed, in the Schwarzschild frame, in
terms of the corresponding spatial three-metric $\gamma_{schw}^{ij}$:
\begin{eqnarray}
dM_B&=&\hat\rho \sqrt{\gamma_{schw}}d^3x\nonumber\\
&=&\hat\rho r_s^2\left(1+\frac{Gm_s(r_s)}{r_sc^2}\right)dr_s\sin\theta d\theta d\phi\,.
\end{eqnarray}
The integrand in the quadrupole moment (\ref{defmhat}) is expressed in the
post-Newtonian coordinates, i.e., it is $r^2=r_s^2(1-2V(r_s)/c^{2})$.  The
integrals $\hat{\cal V},\hat\Pi,\hat{\cal M}$ then take the form
\begin{eqnarray}
\hat{\cal V}&=&-3\hat\Pi\nonumber\\
\hat\Pi&=&4\pi\int_0^{R_{NS}}\hat p\left(1+\frac{Gm_s(r_s)}{r_sc^2}\right)r_s^2dr_s\nonumber\\
\hat{\mathcal{M}}&=&\frac{4\pi}{3}\int^{R_{NS}}_{0}\hat\rho\left(1-\frac{2V(r_s)}{c^2}
+\frac{Gm_s(r_s)}{r_sc^2}\right) r^{4}_{s}dr_{s}\,.\nonumber\\
\end{eqnarray}
%%%%%%%%%%%%%%%%%%%%%%%%%%%%%%%%%%%%%%%%%%%%%%%%%%%%%%%%%%%%%%%%%%%%%
\subsection{The post-Newtonian metric}
%%%%%%%%%%%%%%%%%%%%%%%%%%%%%%%%%%%%%%%%%%%%%%%%%%%%%%%%%%%%%%%%%%%%%
To derive the equations describing the orbital motion of the binary and the
tidal tensor, we shall use a $3$ PN metric written in harmonic coordinates
($\{x^\mu=ct,x,y,z\}$):
\begin{widetext}
\begin{eqnarray}
g_{00}&=&-1+\frac{2V}{c^{2}}-\frac{2V^{2}}{c^{4}}+\frac{8}{c^{6}}\left[\hat{X}+
V_{i}V_{i}+\frac{V}{6}\right]+\frac{32}{c^{8}}\left[\hat{T}-\frac{V\hat{X}}{2}
+\hat{R}_{i}V_{i}-\frac{VV_{i}V_{i}}{2}-\frac{V^{4}}{48}\right]+\mathcal{O}(10)\label{g00}\\
g_{0i}&=&-\frac{4}{c^{3}}V_{i}-\frac{8}{c^{5}}\hat{R}_{i}-\frac{16}{c^{7}}\left[\hat{Y}_{i}
+\frac{1}{2}\hat{W}_{ij}V_{j}+\frac{1}{2}V^{2}V_{i}\right]+\mathcal{O}(9)\label{g0i}\\
g_{ij}&=&\delta_{ij}\left[1+\frac{2}{c^{2}}V+\frac{2}{c^{4}}V^{2}+\frac{8}{c^{6}}
\left(\hat{X}+V_{k}V_{k}+\frac{V^{3}}{6}\right)\right]+
\frac{4}{c^{4}}\hat{W}_{ij}+\frac{16}{c^{6}}\left(\hat{Z}_{ij}+\frac{1}{2}V\hat{W}_{ij}
-V_{i}V_{j}\right)+\mathcal{O}(8)\label{gij}\ ,
\end{eqnarray}
\end{widetext}
where the potentials
$V,V_{i},\hat{X},\hat{W}_{ij},\hat{R}_{i}$,$\hat{Y}_{i}$,$\hat{Z}_{ij}$, are
defined in terms of retarded integrals over the source densities
\cite{BFP98,FBA06}.  We stress that the potential $V$ appearing in the metric of
the two-body system (\ref{g00})-(\ref{gij}), is different from the potential
$V$ in Eq.~(\ref{1PNstar}), which is the metric of a single star.  Since these
potentials are written as expansions of powers of $1/c^{n}$, in the following we
shall identify the order of expansion with a superscript index. Thus, $V^{(0)}$
defines the scalar potential of order $0$ in $1/c$, $V^{(2)}$ is the $1/c^{2}$
term and so on.

%%%%%%%%%%%%%%%%%%%%%%%%%%%%%%%%%%%%%%%%%%%%%%%%%%%%%%%%%%%%%%%%%%%%%
\subsection{The orbital motion}\label{orbital}
%%%%%%%%%%%%%%%%%%%%%%%%%%%%%%%%%%%%%%%%%%%%%%%%%%%%%%%%%%%%%%%%%%%%%
Following \cite{BAF11}, we assume that the orbit evolves as a slow adiabatic
inspiral of a quasi-circular orbit, i.e., the energy lost through gravitational
waves is balanced by a change of the total binding energy $E$ of the system
\begin{equation}\label{def:Ebalance}
\frac{dE}{dt}=-\mathcal{F}\ ,
\end{equation}
where $E$ and the GW flux $\mathcal{F}$ are expressed in terms of the PN
variable
\begin{equation}\label{def:PNx}
x=\left(\frac{Gm\omega}{c^{3}}\right)^{2/3}\ ,
\end{equation}
being $\omega=d\phi/dt$ the orbital frequency. Eq.(\ref{def:Ebalance}) yields
\begin{equation}\label{def:dxdt}
\frac{dx}{dt}=-\frac{\mathcal{F}}{dE/dx}\ .
\end{equation}
We neglect the orbital eccentricity because, due to gravitational wave emission,
the orbit circularizes well before the latest stages of the inspiral which we
are studying \cite{P64}.  We use the approach named ``Taylor T4 approximant''
\cite{DIST01,SOA10}, in which the right-hand side of eq.(\ref{def:dxdt}) is
expanded to $3.5$ PN order including spin terms. We also include the effects of
the NS tidal deformation on the orbital motion, up to next-to-leading-order
\cite{VFH11}. The orbital phase $\phi(t)$ and the orbital frequency $\omega$
are computed by numerically integrating the following ODEs
\begin{eqnarray}
\frac{dx}{dt}&=&\frac{dx}{dt}\bigg\vert_{pp}+\frac{dx}{dt}\bigg\vert_{tidal}\label{def:TaylorT4a}\\
\frac{d\phi}{dt}&=&\frac{c^{3}}{Gm}x^{3/2}\label{def:TaylorT4b}\,.
\end{eqnarray}
where the point-particle contribution reads
\begin{equation}\label{def:TaylorT4pp}
\frac{dx}{dt}\bigg\vert_{pp}=\frac{64}{5}\frac{\nu}{m} x^{5}\sum_{k=0}^{7}a_{k}x^{k/2}
\end{equation}
with the coefficient $a_{k}$ given in Appendix \ref{appa}, and the tidal term is given by
{\small
\begin{eqnarray}\label{def:TaylorT4tid}
\frac{dx}{dt}\bigg\vert_{tid}&=&\frac{32m_{1}\lambda_{2}}{5m^{7}}\left\{12\left[1
+11\frac{m_{1}}{m}\right]x^{10}+\left[\frac{4421}{28}-\frac{12263}{28}\frac{m_{2}}{m}\right.\right.\nonumber\\
&&\left.\left.+\frac{1893}{2}\frac{m_{2}^{2}}{m^{2}}-661\frac{m_{2}^{3}}{m^{3}}\right]x^{11}\right\}+1\leftrightarrow2\ ,
\end{eqnarray}}
where $\lambda_{2}$ is the Love number of the body $2$, and $1\leftrightarrow2$
means the same terms but whit the label $1$ and $2$ exchanged. As we discuss in
Section \ref{seclove}, the values of the Love number for different stellar
models can be computed with our approach, and agree with the values obtained in
the literature \cite{H08}.

The orbital separation $r_{12}$ is evaluated through the PN expression for
$\gamma={Gm}/{r_{12}c^{2}}$, which is known up to order $3$ PN, including
spin terms \cite{Fav11}, and is found solving the equation
\begin{eqnarray}
\frac{d\gamma}{dt}&=&\frac{dx}{dt}\left\{1+2x\left(1-\frac{\nu}{3}\right)+
\frac{5}{2}x^{3/2}\left(\frac{5}{3}s_\ell+\delta\sigma_\ell\right) + \right.\nonumber\\
&+& 3x^{2}\left(1-\frac{65}{12}\nu\right)+\frac{7}{2}x^{5/2}
\left[\left(\frac{10}{3}+\frac{8}{9}\nu\right)s_\ell + \right.\nonumber\\
&+&2\delta\sigma_\ell \left.\right]+4x^{3}\left[1+\left(-\frac{2203}{2520}
-\frac{41}{192}\pi^{2}\right)\nu    + \right. \nonumber\\
&+&\left.\left.\frac{229}{36}\nu^{2}+\frac{\nu^{2}}{81}\right]\right\},\label{dgammadt}
\end{eqnarray}
where $\delta=\frac{m_{1}-m_{2}}{m}$ and the spin variables are defined as follows
\begin{eqnarray}
\textbf{s}_\ell=\frac{c}{G}\frac{\textbf{S}}{m^{2}}&=&\frac{c}{G}
\frac{\textbf{S}_{1}+\textbf{S}_{2}}{m^{2}}\\
{\bf \sigma}_\ell=\frac{c}{G}\frac{{\bf \Sigma}}{m^{2}}&=&\frac{c}{Gm}
\left[\frac{\textbf{S}_{2}}{m_{2}}-\frac{\textbf{S}_{1}}{m_{1}}\right]\,;
\end{eqnarray}
${\bf S}_{i}=(G/c)m_{i}^{2}\tilde{a}_{i}\hat{{\bf s}}_{i}$ are the spin angular
momenta of bodies $i=1,2$, with dimensionless spin parameters $\tilde{a}_{i}$ and
unit direction vectors $\hat{{\bf s}}_{i}$.

It is important to remark that the adiabatic inspiral of the orbital motion and
the ``adiabatic approximation'' for the Love number, are two different
approximations: the first assumes that the orbital timescale is much larger than
that associated to the gravitational wave energy loss ({\it orbital adiabatic
  approximation}); the second assumes, as mentioned in the Introduction and in
Section \ref{seclove}, that the orbital timescale is much larger than the
timescale associated to the NS internal dynamics ({\it Love number adiabatic
  approximation}).  In this paper, we use the orbital adiabatic approximation,
but we drop the Love number adiabatic approximation.
%%%%%%%%%%%%%%%%%%%%%%%%%%%%%%%%%%%%%%%%%%%%%%%%%%%%%%%%%%%%%%%%%%%%%
\subsection{Post-Newtonian tidal deformations}
%%%%%%%%%%%%%%%%%%%%%%%%%%%%%%%%%%%%%%%%%%%%%%%%%%%%%%%%%%%%%%%%%%%%%
Tidal interactions in binary systems have been studied by many authors in the
framework of general relativity (see for instance \cite{M75,M83}). They are
described by the equation of geodesic deviation:
\begin{equation}
\frac{D^{2}\xi^{\alpha}}{D\tau^{2}}+
R^{\alpha}_{~\beta\gamma\delta}u^{\beta}u^{\gamma}\xi^{\delta}=0~,
\label{gdev}
\end{equation}
where $R^{\alpha}_{~\beta\gamma\delta}$ is the Riemann tensor,
$D/D\tau=u^\mu\nabla_\mu$, and, in the present case, $u^{\beta}$ is the
$4-$velocity of the star center $O^*$ and $\xi^{\alpha}$ is the separation
$4-$vector between $O^*$ and a generic fluid element.  By introducing an
orthonormal tetrad field $\{e^\mu_{(i)}\}$ $(i=0,\dots,3)$ associated with the
frame centered in $O^*$, parallel transported along its motion, and such that
$e^{\mu}_{(0)}=u^{\mu}$, Eq.~(\ref{gdev}) can be cast in the form \cite{P56}
\begin{equation}\label{tidaleq}
\frac{d^{2}\xi^{(i)}}{d\tau^{2}}+C^{(i)}_{~(j)}\xi^{(j)}=0\ ,
\end{equation}  
where the $\xi^{(i)}=e^{(i)}_\mu\xi^\mu$, and $C^{(i)}_{~(j)}$ are the
components of the relativistic tidal tensor, defined in Eq.~(\ref{defC}).  In
the affine approach, Eq.~(\ref{tidaleq}) applies with $\xi^{(i)}$ replaced by
$a_i$.

In the following subsections, starting from the $3$ PN metric given in
Eqns.~(\ref{g00})-(\ref{gij}), we write the explicit expression of the parallel
transported tetrad, and compute the tidal tensor assuming equatorial motion.
%%%%%%%%%%%%%%%%%%%%%%%%%%%%%%%%%%%%%%%%%%%%%%%%%%%%%%%%%%%%%%%%%%%%%
\subsubsection{The parallel transported tetrad}
%%%%%%%%%%%%%%%%%%%%%%%%%%%%%%%%%%%%%%%%%%%%%%%%%%%%%%%%%%%%%%%%%%%%%
The orthonormal tetrad associated to the PN metric (\ref{g00})-(\ref{gij}),
satisfies the Fermi-Walker transport equations (expressed in terms of the
coordinate time $t$, rather than of the proper time of $O^*$) \cite{F88}:
\begin{equation}\label{FW}
\frac{de^{\mu}_{(\alpha)}}{dt}=\Pi^{\mu}_{\nu}e^{\nu}_{(\alpha)}\,,
\end{equation}
where 
\begin{equation}
\Pi^{\mu}_{\nu}=-\Gamma^{\mu}_{\nu\lambda}v^{\lambda}
-\frac{1}{c^2}g_{\nu\lambda}\left(a^{\mu}v^{\lambda}-a^{\lambda}v^{\mu}\right)\,,
\end{equation}
$a^{\mu}$ is the $4-$acceleration of $O^*$ and $v^{\mu}=dx^\mu/dt,~\mu=0,3$ its
coordinate velocity, with $v^{0}=c$.  The tetrad vectors are \cite{F88}:
\begin{eqnarray}
e^{t}_{(t)}&=&{\tilde e}^{t}_{(t)}\nonumber \\
e^{j}_{(t)}&=&{\tilde e}^{j}_{(t)}\nonumber\\
e^{t}_{(j)}&=&{\tilde e}^{t}_{(j)}\nonumber\\ 
e^{j}_{(x)}&=&\cos\chi{\tilde e}^j_{(x)}+\sin\chi{\tilde e}^j_{(y)}\nonumber\\ 
e^{j}_{(y)}&=&-\sin\chi{\tilde e}^j_{(x)}+\cos\chi{\tilde e}^j_{(y)}\nonumber\\ 
e^{j}_{(z)}&=&{\tilde e}^j_{(z)}\nonumber\\
\nonumber
\end{eqnarray} 
where 
\begin{eqnarray}\label{tet} 
{\tilde e}^{t}_{(t)}&=&1+\frac{1}{c^{2}}\left[V+\frac{v^{2}}{2}\right]
+\mathcal{O}(4)\\
\nonumber 
{\tilde e}^{j}_{(t)}&=&\frac{v^{j}}{c}+\left[V+\frac{v^{2}}{2}\right]
\frac{v^{j}}{c^{3}}+\mathcal{O}(5)\\
\nonumber 
{\tilde e}^{t}_{(j)}&=&\frac{v^{j}}{c}+\mathcal{O}(3)
\nonumber\\ 
{\tilde e}^{j}_{(k)}&=&\delta^{j}_{k}\left[1-\frac{V}{c^{2}}\right]+\frac{v^{j}v^{k}}{2c^{2}}
+\mathcal{O}(4)\ ,
\nonumber
\end{eqnarray} 
and 
\begin{equation}
\chi=\frac{1}{c^2}Q_{xy}
\end{equation}
is the angle describing geodesic precession and frame dragging, given in terms
of the antisymmetric matrix $\textbf{Q}$ defined as
\begin{equation}
\textbf{Q}(t,t_{0})=\int^{t}_{t_{0}}\left[\textbf{v}\times{(\nabla V-\textbf{a})}
-\nabla\times(V\textbf{v}-2{\textbf{V}})\right]dt\,.
\end{equation} 
$\textbf{V}=\{V_i\}$ is the post Newtonian potential, associated with the
components $g_{0i}$ of the PN metric (\ref{g0i}), and $t_0$ is an arbitrary
integration constant.

Eqns.~(\ref{tet}) reduce to those given in ref.~\cite{F88} with the
identification $V^2=-\psi$, $\gamma \delta_{ij} V =\chi_{ij}$ and
$V_i=-\frac{1}{4}g_i$.
%%%%%%%%%%%%%%%%%%%%%%%%%%%%%%%%%%%%%%%%%%%%%%%%%%%%%%%%%%%%%%%%%%%%%
\subsubsection{The tidal tensor}\label{sec:tidtens}
%%%%%%%%%%%%%%%%%%%%%%%%%%%%%%%%%%%%%%%%%%%%%%%%%%%%%%%%%%%%%%%%%%%%%
Having defined the tetrad field, we have explicitly computed (with the help of
the symbolic manipulation software {\tt maple} and the package {\tt GRTensor})
the Riemann and the tidal tensors, up to order $1/c^{3}$.  The general structure
of the tidal tensor components in terms of the derivatives of the PN potentials,
is given in Appendix \ref{tidalcij}; here we show, as an example, the component
$C_{(x)(x)}$:
\begin{widetext}
{\small
\begin{eqnarray}
C_{(x)(x)}&=&-\pxx\vo+\frac{1}{c^{2}}\Bigg\{-4\pxt\vx+4v^{y}
\pxx\vy-4 v^{y}\pxy\vx-(\py\vo)^{2}-\pxx\voo-\bigg[\ptt +(v^{y})^{2}
(\pyy+2\pxx)\ +\nonumber\\ 
&+&2v^{y}\pyt-
v^{x}v^{y}\pxy\bigg]\vo+2\pxx\vo\vo+
2(\px\vo)^{2}\Bigg\}-\frac{4}{c^{3}}\left\{(\pxt+v^{y}\pxy)\vxx-v^{y}\pxx\vyy+\frac{1}{4}\partial^{2}_{xx}V^{(3)}\right\}
\label{tidxx}
\end{eqnarray}}
\end{widetext} 
where \cite{BFP98,FBA06}
\begin{eqnarray}
\label{PNpot}
V^{(0)}&=&\frac{Gm_{1}}{r_{1}}+1\leftrightarrow2\\
\nonumber
V^{(2)}&=&Gm_{1}\left[Gm_{2}\left(-\frac{r_{1}}{4r^{3}_{12}}-
\frac{5}{4r_{1}r_{12}}+\frac{r^{2}_{2}}{4r_{1}r^{3}_{12}}  \right)\ +\right.\nonumber\\
&-&\left.\frac{(\textbf{n}_{1}\cdot\textbf{v}_{1})^{2}}{2r_{1}}+
\frac{2v^{2}_{1}}{r_{1}}\right]+1\leftrightarrow2\nonumber\\
V^{(3)}&=&-2G\epsilon_{ijk}v_{1}^{i}S_{1}^{j}\partial_{k}\left(\frac{1}{r_{1}}\right)+1\leftrightarrow2\nonumber\\
V^{(0)}_{i}&=&\frac{Gm_{1}v^{i}_{1}}{r_{1}}+1\leftrightarrow2\nonumber\\
V^{(1)}_{i}&=&-\frac{G}{2}\epsilon_{ijk}S^{j}_{1}\partial_{k}
\left(\frac{1}{r_{1}}\right)+1\leftrightarrow2\,.
\nonumber
\end{eqnarray}
To hereafter we omit the parentheses to indicate the tetrad components of the
tidal tensor.  In Eqns.~(\ref{PNpot}) $1\leftrightarrow2$ means the same term
but with the labels $1$ and $2$ exchanged;
$r_{1}=\vert\textbf{x}-\textbf{y}_{1}\vert$ and
$\textbf{n}_{1}=(\textbf{x}-\textbf{y}_{1})/r_{1}$, where $\textbf{x}$ is the
field point and $\textbf{y}_{1}(t)$ the trajectory of $m_{1}$ (and similarly for
$m_{2}$) ; $\textbf{v}_{1}=d\textbf{y}_{1}(t)/dt$ is the coordinate velocity,
$\textbf{r}_{12}=\textbf{r}_{1}-\textbf{r}_{2}$ the relative displacement
between the two masses, and $\textbf{v}_{12}=\textbf{v}_{1}-\textbf{v}_{2}$ the
relative velocity. $\epsilon_{ijk}$ is the $3-$dimensional antisymmetric
Levi-Civita symbol with $\epsilon_{123}=1$, and $S_{1}^{j}$ is the spin-vector.

The steps needed to evaluate the tidal tensor at the center of the NS (i.e., at
location 1 in our conventions) are the following:
\begin{enumerate}
\item estimate the various derivatives of the PN potentials $V,V_{i}$;
\item compute the tidal tensor at the source location
  $[C^{i}_{j}]_{1}=C^{i}_{j}(\textbf{x}\rightarrow \textbf{y}_{1})$, and apply a
  regularization procedure;
\item express all quantities in the two-body center of mass frame;
\item switch to the star principal frame defined in Section~\ref{sec:affine}.
\end{enumerate}
The second point requires further clarifications.  Since the PN metric refers to
pointlike sources, the PN potentials (\ref{PNpot}) diverge when computed at the
source locations $\textbf{x}\rightarrow \textbf{y}_{1}$ and
$\textbf{x}\rightarrow \textbf{y}_{2}$.  To remove this divergence, we apply the
Hadamard regularization procedure \cite{BFP98}, which we briefly describe. Let
$F(\textbf{x},\textbf{y}_{1},\textbf{y}_{2})$ be a function depending on the
field point $\textbf{x}$ and on the two source locations
$\textbf{y}_{1}$,$\textbf{y}_{2}$, and admitting, when $\textbf{x}$ approaches
$\textbf{y}_{1}$, an expansion of the form
\begin{equation}\label{exhad}
  F(\textbf{x},\textbf{y}_{1},\textbf{y}_{2})=\sum_{k}r_{1}^{k}
  f_{k}(\textbf{n}_{1},\textbf{y}_{1},\textbf{y}_{2})  \qquad k\in\mathbb{Z}\,.
\end{equation}
The regularized value of $F$ at the point $1$ is the \emph{Hadamard part finie},
which is the average, with respect to the direction $\textbf{n}_{1}$, of the
$k=0$ term in the sum (\ref{exhad}):
\begin{equation}
(F)_{1} =F(\textbf{y}_{1},\textbf{y}_{1},\textbf{y}_{2})=\int
\frac{d\Omega(\textbf{n}_{1})}{4\pi}f_{0}(\textbf{n}_{1},\textbf{y}_{1},\textbf{y}_{2})\,.
\end{equation}
We use this procedure to evaluate $(V)_{1},(V_{i})_{1}$ and their derivatives at
the source point. We remark that the regularization procedure should be applied
separately on the Riemann tensor and on the orthonormal tetrad. However, at the
PN order we are considering, it is perfectly equivalent to apply the
regularization procedure directly to the tidal tensor
$C_{(i)(j)}=R_{\alpha\beta\gamma\delta}e^{\alpha}_{(0)}e^{\beta}_{(i)}e^{\gamma}_{(0)}e^{\delta}_{(j)}$.

We now express the point particle positions $\textbf{y}_{1,2}$ in the system
center of mass frame, by the following coordinate transformation \cite{BI03}
\begin{eqnarray}
y_{1}^{i}&=&\left[\frac{m_{2}}{m}+\nu\frac{m_{1}-m_{2}}{m}\mathcal{P}\right]
r_{12}^{i}+\mathcal{O}(4)\ ,\\
y_{2}^{i}&=&\left[-\frac{m_{1}}{m}+\nu\frac{m_{1}-m_{2}}{m}\mathcal{P}\right]
r_{12}^{i}+\mathcal{O}(4)~,
\nonumber
\end{eqnarray}
where
\begin{equation}
\mathcal{P}=\frac{1}{c^{2}}\left[\frac{v_{12}^{2}}{2}-
\frac{Gm}{2r_{12}}\right]+\mathcal{O}(4)\ .
\end{equation}
Finally, we express $C^{i}_{j}$ in the principal frame using the rotation matrix
\begin{equation}
 T=\left(\begin{array}{ccc}
\cos\psi & \sin\psi &0\\
-\sin\psi & \cos\psi & 0\\
0 & 0 &1    
\end{array}\right)
\end{equation}
 where $\psi$ is defined by Eq.~(\ref{NSspin})
\begin{equation}
\frac{d\psi}{d\tau}=\Omega\,.
\end{equation}
The complete form of the tidal tensor $c=TCT^{T}$ is given by:
\begin{widetext}
{\small
\begin{eqnarray}
c_{xx}=&-&\frac{Gm_2}{2r_{12}^3}\left\{1+3\cos[2\psi_{l}]\right\}+
\frac{G}{4c^{2}r_{12}^4}\bigg\{\left[6Gm_{2}^2+5G m \mu+3 \dot{r}_{12}^2
m_{1} \nu r_{12}
\right]\left(1+3\cos[2\psi_{l}]\right)-6\dot{\phi}^2m_{2}r_{12}^3
(1+\cos[2\psi_{l}])\ +\nonumber\\
&&\phantom{a}\nonumber\\
&+&6m_{2}r^{2}_{12}\left(\frac{m_{2}^{2}}{m^{2}}+2\nu\right)\dot{\phi}
\dot{r}_{12}
 \sin[2\psi_{l}]\bigg\}+\frac{3GS_{2}^{z}}{mc^{3}r_{12}^{3}}\bigg\{\dot{\phi} 
(m_{2}-m_{1})+(m_{2}-5m_{1})\dot{\phi}\cos[2\psi_{l}]+\dot{r}_{12}
\frac{(m_{2}+3m_{1})}{r_{12}}
\sin[2\psi_{l}]\bigg\}\label{cxx}\\
&&\phantom{a}\nonumber\\
c_{yy}=&-&\frac{Gm_{2}}{2r_{12}^3}\left\{1-3\cos[2\psi_{l}]\right\}+
\frac{G}{4 c^2 r_{12}^4}\bigg\{\left[6Gm_{2}^2+5G m \mu+3 \dot{r}_{12}^2 m_{1} 
\nu r_{12}\right]\left(1-3\cos[2\psi_{l}]\right)-6\dot{\phi}^2
m_{2}r_{12}^3(1-\cos[2\psi_{l}])\ +\nonumber\\
&&\phantom{a}\nonumber\\
&-&6m_{2}r^{2}_{12}
\left(\frac{m_{2}^{2}}{m^{2}}+2\nu\right)\dot{\phi} 
\dot{r}_{12} \sin[2\psi_{l}]\bigg\}+\frac{3
S_{2}^{z}}{mc^{3} r_{12}^3}\bigg\{\dot{\phi}
 (m_{2}-m_{2})-\dot{\phi} (m_{2}-5m_{1}) \cos[2\psi_{l}]-\dot{r}_{12} \frac{(m_{2}
+3m_{1})}{r_{12}}\sin[2\psi_{l}]\bigg\}\label{cyy}\\
&&\phantom{a}\nonumber\\
c_{zz}&=&\frac{Gm_{2}}{r_{12}^3}-\frac{G}{c^2}\left[3\frac{Gm_{2}^2}{r_{12}^{4}}
+\frac{5}{2}\frac{Gm\mu}{r_{12}^4}+\frac{3}{2}\frac{m_{1}\nu\dot{r}_{12}^2}{r_{12}^{3}}-
\frac{3}{r_{12}}m_{2}\dot{\phi}^2
\right]-\frac{6G(m_{2}-m_{1})S_{2}^{z}}{mc^{3}r_{12}^3}\dot{\phi}\label{czz}\\
&&\phantom{a}\nonumber\\
c_{xy}&=&\frac{3Gm_{2}}{2r_{12}^{3}}\sin[2\psi_{l}]+\frac{3G}{4 c^2 r_{12}^4}
\bigg\{2m_{2}r^{2}_{12}\left(\frac{m_{2}^{2}}{m^{2}}+2\nu\right)
\dot{\phi}\dot{r}_{12}
 \cos[2\psi_{l}]-\left[6G m_{2}^2+5G m \mu+3 \dot{r}_{12}^2 m_{1} 
\nu r_{12}- 2 \dot{\phi}^2 m_{2} r_{12}^3\right]\ \times\nonumber\\
&&\phantom{a}\nonumber\\
&\times& \sin[2\psi_{l}]\bigg\}-\frac{3G S_{2}^{z}}{mc^{3}
r_{12}^4}\left\{\dot{r}_{12}
 (m_{2}+3 m_{1}) \cos[2\psi_{l}]+\dot{\phi} (m_{2}-5m_{1}) r_{12} 
\sin[2\psi_{l}]\right\}\label{cxy}
\end{eqnarray}}
\end{widetext}
where the \emph{lag} angle $\psi_{l}=\psi-\phi+\chi$ describes the misalignment
between the axis $a_{1}$ and the line between the two objects. In the tidal
tensor components the dot indicates differentiation with respect to the
coordinate time $t$.  In the principal frame, the geodesic deviation equation
for the tidal deformation can be written as
\begin{equation}
\frac{d^2a_i}{d\tau^2}+c_{ij}a_j=0\,.\label{devgeodprinc}
\end{equation}

It should be stressed that, as noted in \cite{B06}, if the system is in
quasi-circular inspiral, the radial motion is due only to gravitational
back-reaction; consequently $\dot{r}_{12}\simeq ({\bf n}_{12}{\bf v}_{12})\sim
1/c^5$ and can be neglected.

%%%%%%%%%%%%%%%%%%%%%%%%%%%%%%%%%%%%%%%%%%%%%%%%%%%%%%%%%%%%%%%%%%%%%
\subsubsection{Comparison with previous expressions of the tidal tensor}
%%%%%%%%%%%%%%%%%%%%%%%%%%%%%%%%%%%%%%%%%%%%%%%%%%%%%%%%%%%%%%%%%%%%%
As a first check, we compare the tidal tensor derived in \cite{M83} for a test
particle ($\nu\rightarrow 0$) moving along a geodesic, with our tidal
tensor. This tensor has been used in the literature to study tidal effects in
binary systems using a quasi-stationary approach \cite{WL00,FGP09}, or evolving
the orbital equations, assuming quasi-circular orbit \cite{PTR11}.  Let us
consider, as an example, the $c_{xx}$ component for a non-rotating BH.  Eq.~(70)
of ref. \cite{M83} gives
\begin{equation}\label{cxxScw}
c_{xx}^{sch}=\frac{Gm_{2}}{r_{s}^{3}}\left(1-
3\frac{r^{2}_{s}+K}{r^{2}_{s}}\cos[\psi_{l}]^{2}\right)\ ,
\end{equation}
where 
\begin{equation}
K=\frac{L_{z}^{2}}{c^2}= \frac{1}{c^2}\left(\frac{d\phi}{d\tau}\right)^2r^4
=\frac{1}{c^2}\left(\frac{d\phi}{dt}\right)^2r^4+O\left(\frac{1}{c^4}\right)\,,\label{defK}
\end{equation}
and $r_{s}$ is the radial distance in Schwarzschild coordinates. In order to
compare Eq.~(\ref{cxxScw}) with Eq.~(\ref{cxx}) we need to express
$c_{xx}^{sch}$ in terms of the same radial coordinate adopted for the PN
expansion
\begin{equation}\label{R-r}
r_{s}=r_{12}\left(1+\frac{Gm_{2}}{2c^{2}r_{12}}\right)^{2}\,.
\end{equation}
We find (up to $1/c^3$ terms)
\begin{eqnarray}
c_{xx}^{sch}&=&-\frac{Gm_{2}}{2r_{12}^{3}}\left(1+3\cos[2\psi_{l}]\right)
+\frac{3Gm_{2}}{2c^2r_{12}^{4}}\left\{Gm_{2}\ +\right. \nonumber\\
&+&\left.3Gm_{2}\cos[2\psi_{l}]- \dot{\phi}^2r^3_{12}- 
\dot{\phi}^2r^3_{12}\cos[2\psi_{l}]\right\}\,.
\end{eqnarray}
This expression coincides with our Eq.~(\ref{cxx}), in the limit $\nu\rightarrow
0$ and $\dot{r}_{12}\simeq 0$.

We would like to make a further remark about the difference between the tidal
tensor (\ref{cxxScw}), derived from the Schwarzschild metric assuming that
$m_{1}$ follows a time-like geodesic of the Schwarzschild spacetime, and that
derived from a two-body post-Newtonian metric.  For a particle in circular
orbit, the constant $K$ given in Eq.~(\ref{defK}) is
\begin{equation}
\frac{K}{r^{2}_{s}}=\frac{Gm_{2}}{r_{s}c^2-3Gm_{2}}\,.
\end{equation}
The former equation diverges for $r_{s}\rightarrow 3Gm_{2}/c^2$.  This divergence is
present also in the tidal tensor components, as shown by Eq.~(\ref{cxxScw}), and
it may affect the evaluation of tidal effects even if the distance between the
interacting bodies is larger than (but close to) $r_s=3Gm_2/c^2$.

Conversely, as stressed in \cite{B06}, such divergence does not appear in the PN
equations of motion, and consequently the tidal tensor components
(\ref{cxx})-(\ref{cxy}) are free of this unphysical behaviour.

On the other hand, as $\nu \rightarrow 0$ our approach loses accuracy, since in
the test particle limit the PN expansion is poorly convergent \cite{B06}.

As a second check we compare our tidal tensor with that used in \cite{VFH11},
previously derived in \cite{DSX92} up to order $\sim1/c^2$ with a completely
different approach, based on a multipole expansion.  Comparing $-G_2^{ij}$
(Eq.~(2.2) of \cite{VFH11}) with our tensor $C_{ij}$, truncated to order
$\sim1/c^2$, we find that (renaming $m_1\leftrightarrow m_2$) they coincide.

%%%%%%%%%%%%%%%%%%%%%%%%%%%%%%%%%%%%%%%%%%%%%%%%%%%%%%%%%%%%%%%%%%%%%
\subsection{Internal dynamics}\label{sec:intdyn}
%%%%%%%%%%%%%%%%%%%%%%%%%%%%%%%%%%%%%%%%%%%%%%%%%%%%%%%%%%%%%%%%%%%%%
The internal dynamics of the NS is described using the Hamiltonian approach in
the affine approximation \cite{CL85,WL00}, recently improved to take into
account general relativistic effects \cite{FGP09}:
\begin{equation}\label{fullhamilt}
\mathcal{H}=\mathcal{H}_{T}+\mathcal{H}_{I}
\end{equation}
where $\mathcal{H}_{I}$ describes the NS internal structure, and
$\mathcal{H}_{T}$ describes the tidal interaction.  $\mathcal{H}_{I}$ is
obtained directly from the internal Lagrangian $\mathcal{L}_I$
(\ref{lagrangeI}). The tidal Hamiltonian $\mathcal{H}_{T}$ is obtained from the
tidal Lagrangian (built up with the coefficients $c_{ij}$,
Eqns.~(\ref{cxx})-(\ref{czz}) ):
\begin{equation}
\mathcal{L}_{T}=-\frac{1}{2}c_{ij}I_{ij}\,,\label{LT}
\end{equation}
where 
\begin{eqnarray}\nonumber
\textbf{I}=\hat{\mathcal{M}}\cdot diag\left(\frac{a_{i}}{R_{NS}}\right)^{2}
\end{eqnarray}
is the inertia tensor of the star in the principal frame. 

In deriving  the dynamical equations from the Hamiltonian
(\ref{fullhamilt}), we use the PN time coordinate $t$, which
is related to the proper time of the star center of mass $\tau$ by the relation
$d\tau=\gamma(t)^{-1} dt$ where the redshift factor $\gamma(t)$ is:
\begin{eqnarray}\label{lorenztg}
\gamma(t)=&1&+\frac{1}{c^{2}}\left(\frac{m_{2}^{2}}{m^{2}}
\frac{v_{12}^{2}}{2}+\frac{Gm_{2}}{r_{12}}\right)\ +\nonumber\\
&+&\frac{1}{8c^{4}r_{12}^{2}}\bigg\{4G^{2}(m_{2}^{2}-3m\mu)
+4Gr_{12}\bigg[-m_{1}\nu\dot{r}_{12}\nonumber\\
&+&\frac{m_{2}}{m^{3}}\left(4m_{1}^{3}+11m_{1}^{2}m_{2}+14m_{1}m_{2}^{2}+5m_{2}^{3}\right)v_{12}^{2}\bigg]\nonumber\\
&+&\frac{m_{2}^{2}}{m^{4}}\left(4m_{1}^{2}-4m\mu+3m_{2}^{2}\right)r_{12}^{2}v_{12}^{4}\bigg\}\ .
\end{eqnarray}
It should be mentioned that in previous works, where the affine approach
including relativistic corrections was used \cite{WL00,FGP09,PTR11}, the
contribution of the redshift factor was neglected, i.e., it was assumed
$t\simeq\tau$.

We also remark that $\mathcal{H}_{T}\simeq-\mathcal{L}_{T}$, since
$\mathcal{L}_{T}$ does not depend on the conjugate momenta.

The equations of motion for the variables
$q_{i}=\{\psi,\lambda,a_{1},a_{2},a_{3}\}$ and their conjugate momenta
$p_{i}=\{p_{\psi},p_{\lambda},p_{a_{1}},p_{a_{2}},p_{a_{3}}\}$ are:
\begin{eqnarray}
  \frac{da_{1}}{dt}&=&\frac{R_{NS}}{\gamma(t)}\frac{p_{a_{1}}}{\hat{\mathcal{M}}}\label{da1}\\
  \frac{da_{2}}{dt}&=&\frac{R_{NS}}{\gamma(t)}\frac{p_{a_{2}}}{\hat{\mathcal{M}}}\label{da2}\\
  \frac{da_{3}}{dt}&=&\frac{R_{NS}}{\gamma(t)}\frac{p_{a_{3}}}{\hat{\mathcal{M}}}\label{da3}\\
  \frac{dp_{a_{1}}}{dt}&=&\frac{\hat{\mathcal{M}}}{\gamma(t)}\Bigg[\Lambda^{2}+
  \Omega^{2}-2\frac{a_{2}}{a_{1}}\Lambda\Omega+\frac{1}{2}
  \frac{\hat{\cal V}}{\hat{\mathcal{M}}}R_{NS}^{3}\tilde{A}_{1}\nonumber\\
  &+&\frac{R_{NS}^{2}}{\hat{\mathcal{M}}}\frac{\Pi}{a_{1}^{2}}-c_{xx}\Bigg]a_{1}\label{eqa1}\\
  \frac{dp_{a_{2}}}{dt}&=&\frac{\hat{\mathcal{M}}}{\gamma(t)}\Bigg[\Lambda^{2}
  +\Omega^{2}-2\frac{a_{1}}{a_{2}}\Lambda\Omega+
  \frac{1}{2}\frac{\hat{\cal V}}{\hat{\mathcal{M}}}R_{NS}^{3}\tilde{A}_{2}\nonumber\\ 
  &+&\frac{R_{NS}^{2}}{\hat{\mathcal{M}}}\frac{\Pi}{a_{2}^{2}}-c_{yy}\Bigg]a_{2}\label{eqa2}\\
  \frac{dp_{a_{3}}}{dt}&=&\frac{\hat{\mathcal{M}}}{\gamma(t)}\left[\frac{1}{2}
    \frac{\hat{\cal V}}{\hat{\mathcal{M}}}R_{NS}^{3}\tilde{A}_{3}
    +\frac{R_{NS}^{2}}{\hat{\mathcal{M}}}\frac{\Pi}{a_{3}^{2}}-c_{zz}\right]a_{3}\label{eqa3}\\
  \frac{d\lambda}{dt}&=&\frac{\Lambda}{\gamma(t)}\\
  \frac{dp_{\lambda}}{dt}&=&\frac{1}{\gamma(t)}\frac{d{\mathcal{C}}}{d\tau}=0\\
  \frac{d\psi}{dt}&=&\frac{\Omega}{\gamma(t)}\\
  \frac{dp_{\psi}}{dt}&=&\frac{1}{\gamma(t)}\frac{dJ^z}{d\tau}=
  \frac{\hat{\mathcal{M}}}{R_{NS}}\frac{c_{xy}}{\gamma(t)}
  \left(a_{2}^{2}-a_{1}^{2}\right)\,,\label{dppsi}
\end{eqnarray}
where 
\begin{equation}
\tilde A_i\equiv\int_0^\infty\frac{du}{(a_i^2+u)\sqrt{(a_1^2+u) (a_2^2+u) (a_3^2+u)}}
\,,\label{defAi}
\end{equation}
$J^z$ is the NS angular momentum, and $\mathcal{C}$ is 
the conjugate momentum associated to $\lambda$:
\begin{eqnarray}
\mathcal{C}&=&\frac{\hat{\mathcal{M}}}{R_{NS}^{2}}\left[\left(a_{1}^{2}
+a_{2}^{2}\right)\Lambda-2a_{1}a_{2}\Omega\right]\nonumber \\
J^z&=&\frac{\hat{\mathcal{M}}}{R_{NS}^{2}}\left[\left(a_{1}^{2}
+a_{2}^{2}\right)\Omega-2a_{1}a_{2}\Lambda\right]\,.
\end{eqnarray}
${\cal C}$ can be interpreted as the circulation of the fluid \cite{WL00}, i.e., the
line integral of the four-velocity on a closed worldline enclosing the system.
In absence of viscosity, $\mathcal{C}$ is a constant of motion.

It is worth mentioning that, in the spherical configuration ($a_i=R_{NS}$), the
integrals (\ref{defAi}) can be solved analytically, finding $\tilde
A_i=2/(3R_{NS}^3)$; furthermore, the virial theorem (\ref{virial}) implies that
${\hat{\cal V}}=-3\hat\Pi$. Consequently, in the spherical limit the terms in
$\hat{\cal V}$ and $\Pi$ in Eqns.~(\ref{eqa1})-(\ref{eqa3}) cancel:
\begin{equation}
\left[\frac{1}{2}\frac{\hat{\cal V}}{\hat{\mathcal{M}}}R_{NS}^{3}\tilde{A}_{i}
  +\frac{R_{NS}^{2}}{\hat{\mathcal{M}}}\frac{\Pi}{a_{i}^{2}}\right]_{sph}=0\,.
\label{property}
\end{equation}
This property is crucial to ensure a stable evolution.  Indeed, the system
(\ref{da1})-(\ref{dppsi}) admits an equilibrium solution, for which the star is
spherical and non-rotating, and the tidal tensor vanishes, only if the property
(\ref{property}) is satisfied.  If such solution exists, the tidal deformation
induced by the interaction is basically a perturbation of the equilibrium
configuration, and the system of equations is well behaved.  Conversely, if the
cancellation (\ref{property}) is not exact, the system becomes unstable, because
the terms in $\hat{\cal V}$ and $\Pi$ are larger than other terms and the
equations are non-linear. We remark that the validity of Eq.~(\ref{property}) is
guaranteed in our approach, because the virial theorem is satisfied exactly, as
discussed in Section \ref{sec:affine}.
%%%%%%%%%%%%%%%%%%%%%%%%%%%%%%%%%%%%%%%%%%%%%%%%%%%%%%%%%%%%%%%%%%%%%
\subsection{Roche lobe and mass shedding} 
%%%%%%%%%%%%%%%%%%%%%%%%%%%%%%%%%%%%%%%%%%%%%%%%%%%%%%%%%%%%%%%%%%%%%
In the next Section we shall compare the results obtained by numerically
integrating the equations of motion (\ref{da1})-(\ref{dppsi}) for a a BH-NS
coalescence, with those published in the literature; we shall evolve the
equations up to an orbital separation, $r_{shed}$, at which mass shedding sets
in.

In order to find the value of $r_{shed}$, we estimate the Roche lobe radius of
the NS during the inspiral. It defines the region surrounding the star where a
particle of mass $m_{0}\ll m_{1}$ is bounded to the NS gravitational
attraction. Following the strategy adopted in \cite{FDB00}, we estimate the
three-body potential for masses $m_{0}\ll m_{1}\leq m_{2}$ (at Newtonian order)
for equatorial orbits in the $x-y$ plane:
\begin{equation}\label{potU}
U(x,y)=-\frac{Gm_{1}}{\vert\textbf{x}-\textbf{y}_{1}\vert}-
\frac{Gm_{2}}{\vert\textbf{x}-\textbf{y}_{2}\vert}-\frac{1}{2}\omega^{2}x^{2}
\end{equation}  
where $\textbf{y}_{1/2}$ are the $m_{1/2}$ displacement vectors, and the last
term is the centrifugal contribution with
\begin{eqnarray}
\omega=\sqrt{\frac{G(m_{1}+m_{2})}{r_{12}^{3}}}\nonumber \ .
\end{eqnarray}
Since $U(x,y)$ takes its maximum, $U_{Rl}$, on the surface defining the Roche
lobe, we compute numerically $U(x,y)$, finding the Roche lobe on the $x-y$
plane. Mass shedding starts when the star, which is stretched along the direction
of the axis $a_1$, touches the Roche lobe.

We also determine the location of the $3$ PN ICO (Innermost Circular Orbit),
$r_{ICO}$; to this aim we minimize the total binding energy of the BH-NS system,
including the spin contribution \cite{Fav11}.  If $r_{shed}> r_{ICO}$, the NS is
disrupted before the merger.

It should be noted that the affine approach is intrinsically non-linear,
therefore it takes into account non-linear hydrodynamical effects. It also can
describe mode oscillations (see, for instance, \cite{CFS06} and, in a similar
framework, \cite{L94}). However, it does not account for non-linearities in
the tidal tensor; since these effects are of the order of $(R_{NS}/r_{12})^5$ \cite{FH08},
which never exceed $\sim10^{-4}$, they can be safely discarded.

Higher multipoles (octupole, etc.) of the tidal field are also neglected in the
affine approach; they are suppressed by a factor $\sim (R_{NS}/r_{12})^2$
\cite{MW04,FH08}; higher PN orders are suppressed by a factor $\sim
m/(r_{12}c^2)$. Both these quantities can become as large as $\sim0.15$ as
$r_{12}\rightarrow r_{shed}$. Therefore, our approach becomes less accurate in
the last stages of the inspiral.  We plan to improve our model, computing higher
PN orders and higher multipole contributions to the tidal field, in future
publications.
%%%%%%%%%%%%%%%%%%%%%%%%%%%%%%%%%%%%%%%%%%%%%%%%%%%%%%%%%%%%%%%%%%%%
\section{Dynamical tests}
\label{validate}
%%%%%%%%%%%%%%%%%%%%%%%%%%%%%%%%%%%%%%%%%%%%%%%%%%%%%%%%%%%%%%%%%%%%  
To validate our approach, we have integrated the dynamical equations
(\ref{def:TaylorT4a})-(\ref{dgammadt}) and (\ref{da1})-(\ref{dppsi}) to simulate
BH-NS binary coalescences. In order to compare our results with the existing
literature, we assume that the neutron star is irrotational (i.e., we set
$\mathcal{C}=0$), while the black hole can rotate.
%%%%%%%%%%%%%%%%%%%%%%%%%%%%%%%%%%%%%%%%%%%%%%%%%%%%%%%%%%%%%%%%%%%%
\subsection{Fully relativistic simulations}
%%%%%%%%%%%%%%%%%%%%%%%%%%%%%%%%%%%%%%%%%%%%%%%%%%%%%%%%%%%%%%%%%%%%  

We compare our results with fully relativistic simulations from three groups,
who have kindly shared the required data with us: the Potsdam group
\cite{privcomm} (which we denote by AEI); the Urbana group \cite{E09} (URB);
the Kyoto/Tokyo group \cite{S09} (KT). All simulations (including ours) use the
same $\Gamma=2$ polytropic equation of state. The values of the mass ratio
$q=m_2/m_1=M_{BH}/M_{NS}$, of the BH spin parameter $\tilde a=a/M_{BH}$, and of
the NS compactness $C=M_{NS}/R_{NS}$ are:
\begin{enumerate}
\item 
AEI simulations: $q=5$, $\tilde a=0$,
$C=0.1,0.125,0.15$; 
\item 
URB simulations: $q=3$, $\tilde a=0,0.75$,
$C=0.145$; 
\item
KT simulations: $q=2,3$, $\tilde a=0$, $C=0.145$.
\end{enumerate}

In order to check the validity of our PN formulae, and to determine the time
offset $t_{off}$ between our simulations and the fully relativistic simulations,
as a preliminary check we compare the orbital motion.  It is worth remarking
that the time offset is needed to compare our results with the fully
relativistic simulations.  In particular, it is needed to compare, for each
simulation, the time $t_{shed}$ at which our model predicts the onset of mass
shedding, with the time at which this occurs in the corresponding fully
relativistic simulation. To this aim we follow the startegy adopted in
\cite{BBK07}: we demand that the PN and the numerical gravitational wave
frequencies agree at some fiducial frequency $\omega_{m}$, defining $t_{off}$ as
the time for which $\dot{\phi}_{PN}(t_{off})=\omega_{m}$.

We remark that since the frequency of the $m=2$ component of the gravitational
wave, $\Omega_{GW}=2\Omega$, is a gauge invariant quantity (see for instance
\cite{BBK07,HHGSB08,CLNZ09} and references therein), it is an appropriate
quantity for our comparisons.  On the contrary, it is impossible to directly
compare the radial coordinates, since the gauge used by fully relativistic
simulations is different from our gauge, and furthermore it changes dynamically
during the simulation \cite{HHGSB08}.  Note that here $m$ is the harmonic index,
not the total mass as in the rest of the paper.

\begin{figure*}[t]
\epsfig{file=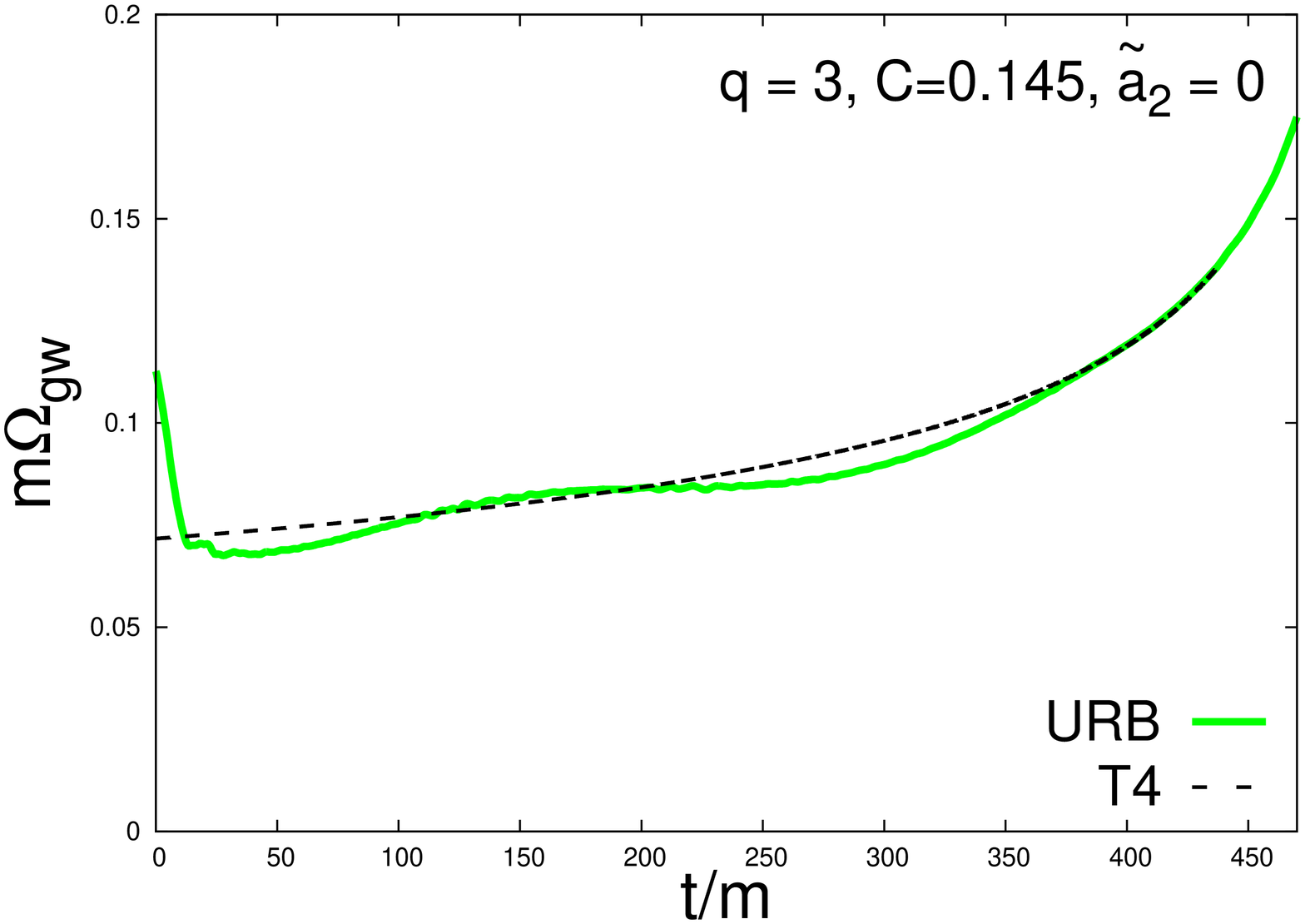,width=130pt,angle=270}
\epsfig{file=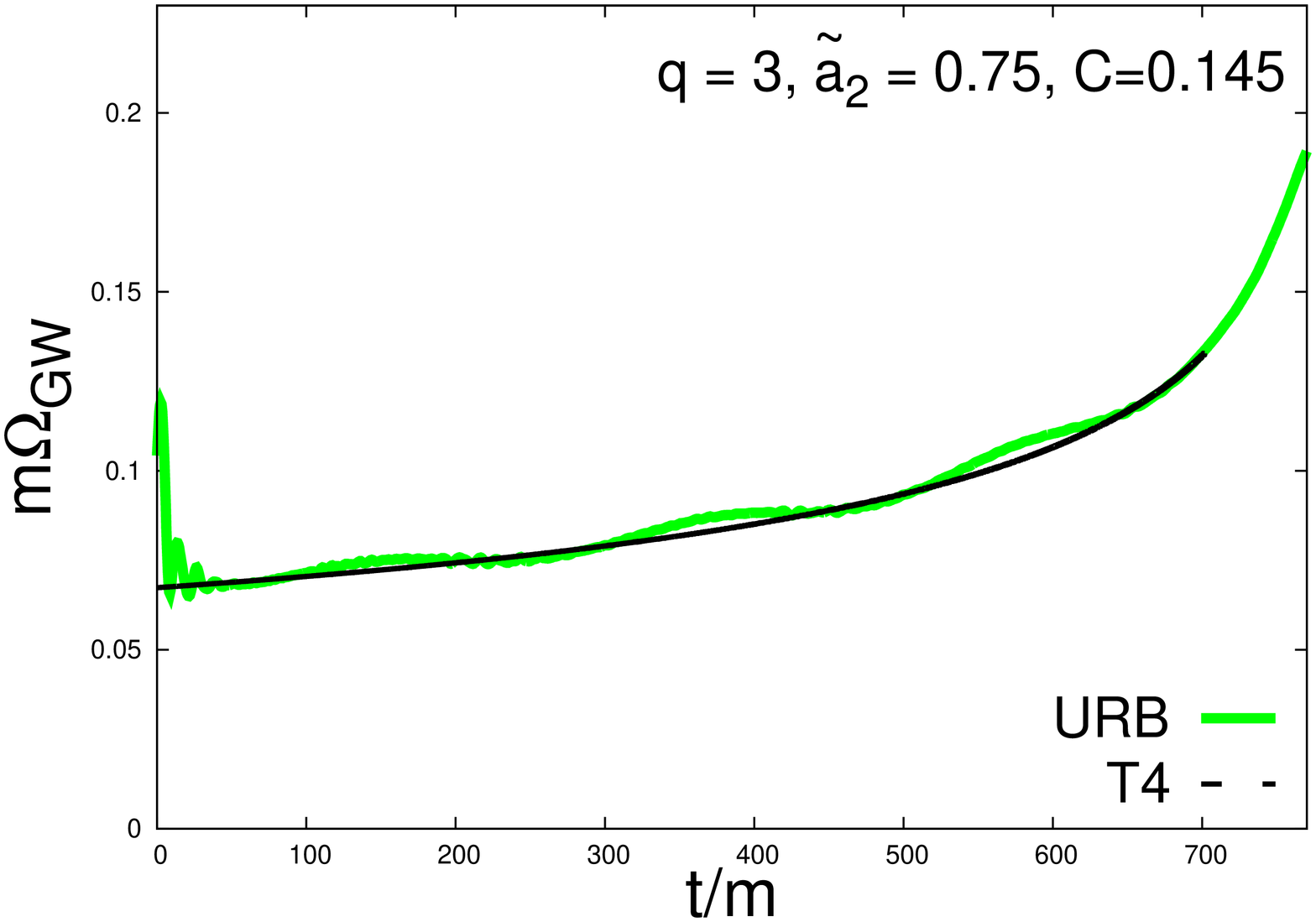,width=130pt,angle=270}
\epsfig{file=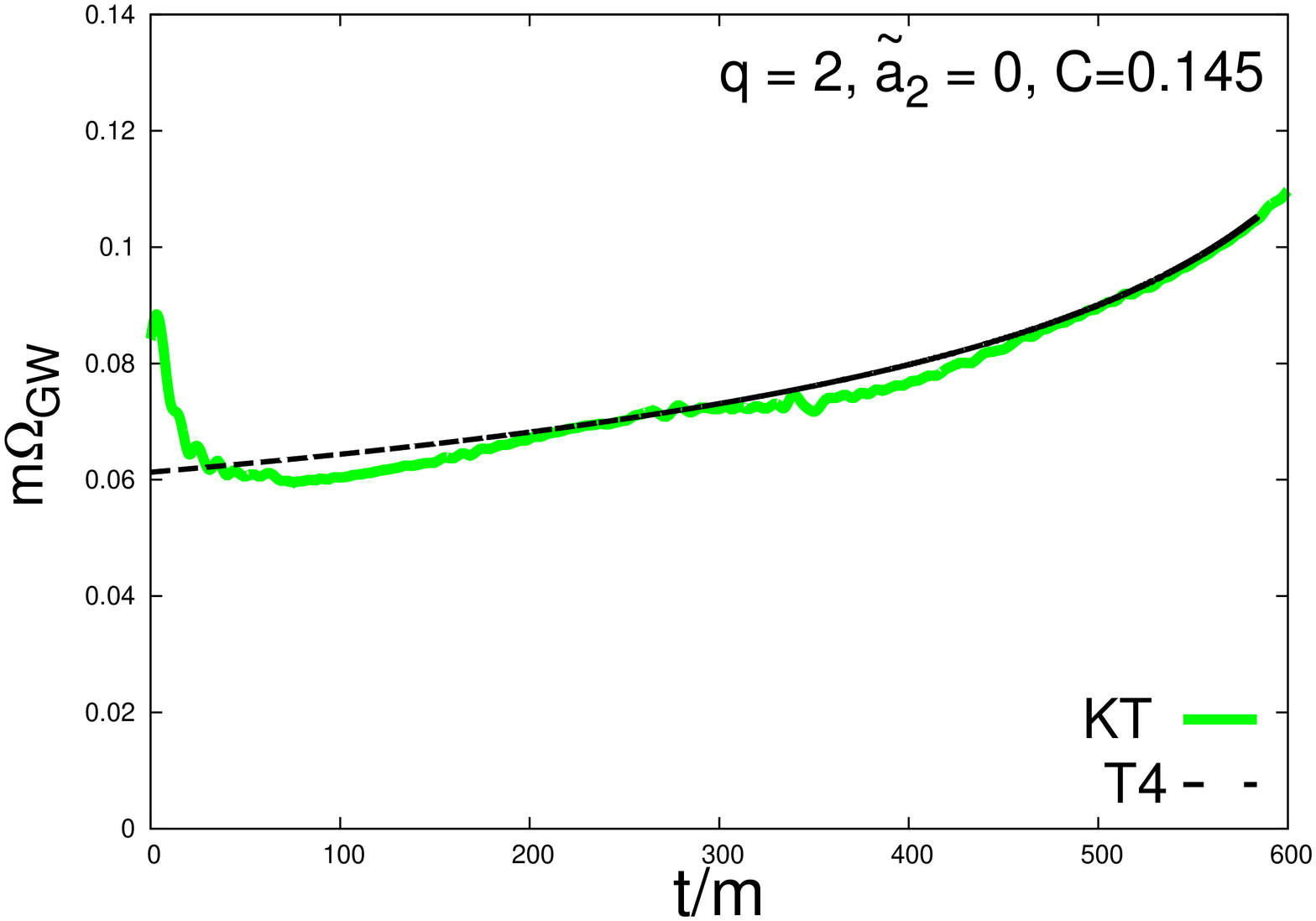,width=130pt,angle=270}
\epsfig{file=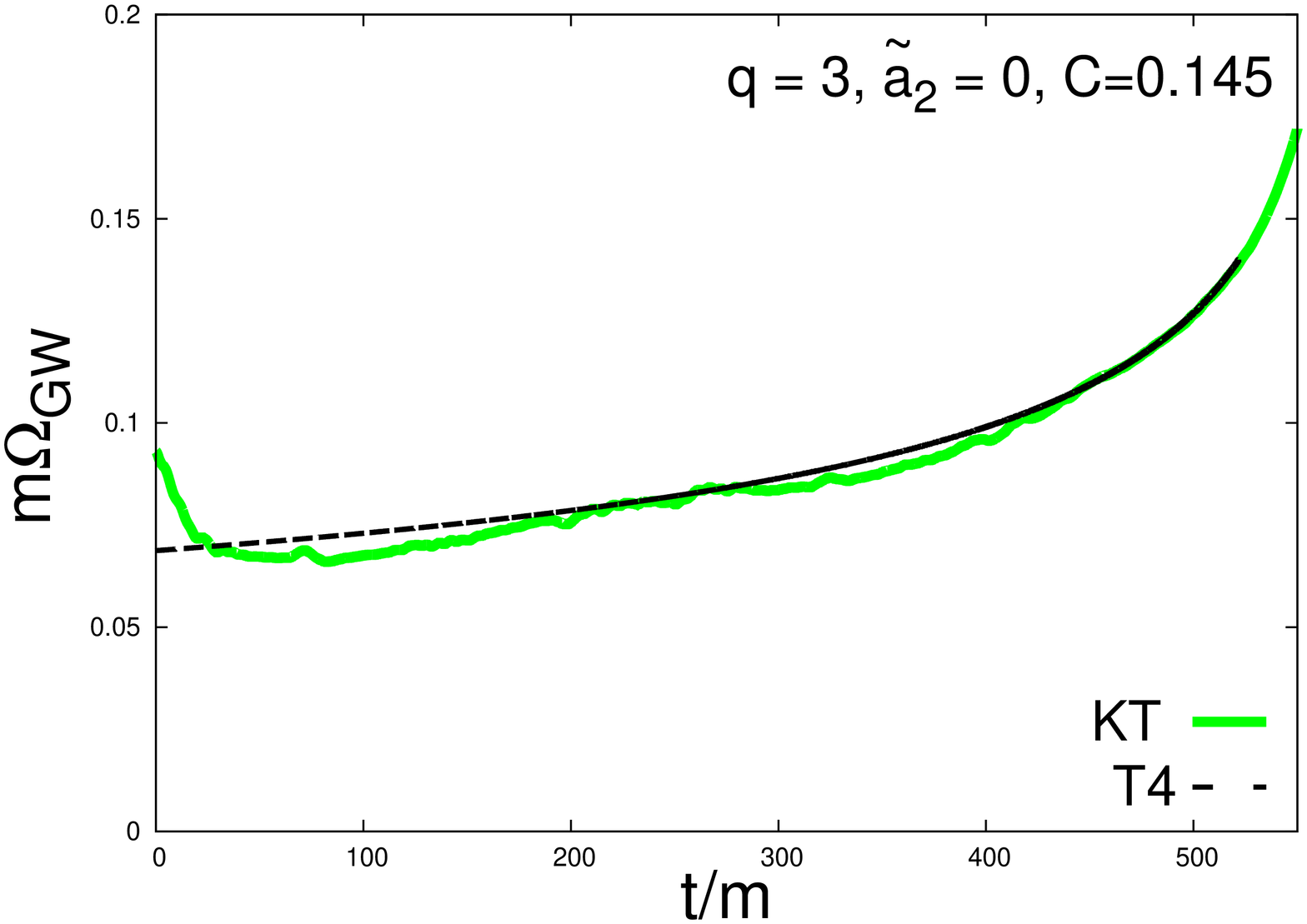,width=130pt,angle=270}
\caption{(Color online) We plot the gravitational wave frequency $\Omega_{GW}$
  versus time, both normalized to the total mass of the binary.  The URB, KT
  data are indicated by a solid line, and our data by a dashed line.  The dashed
  lines stop at $r_{12}=r_{shed}$, where the deformed star touches the Roche
  lobe.}
\label{fig_omega}
\end{figure*}

The comparison with the URB and KT data is shown in Fig.~\ref{fig_omega}, where
we plot $\Omega_{GW}$ versus time (both normalized to the total mass of the
binary).  Our profiles are indicated in Fig.~\ref{fig_omega} with a dashed line,
which ends at the onset of mass-shedding. The URB and KT profiles are shown by a
continuous line.

As expected \cite{BBK07,HHGSB08,CLNZ09,LKSB11,BDGNR11}, the PN (and EOB)
description of the inspiral phase is in good agreement with fully relativistic
simulations. The oscillations in the URB, KT curves shown in
Fig.~\ref{fig_omega}, are due to the fact that their initial data have a
residual eccentricity, while our orbits are quasi-circular. A comparison with
the AEI data gives similar results.

\begin{figure*}[t]
\epsfig{file=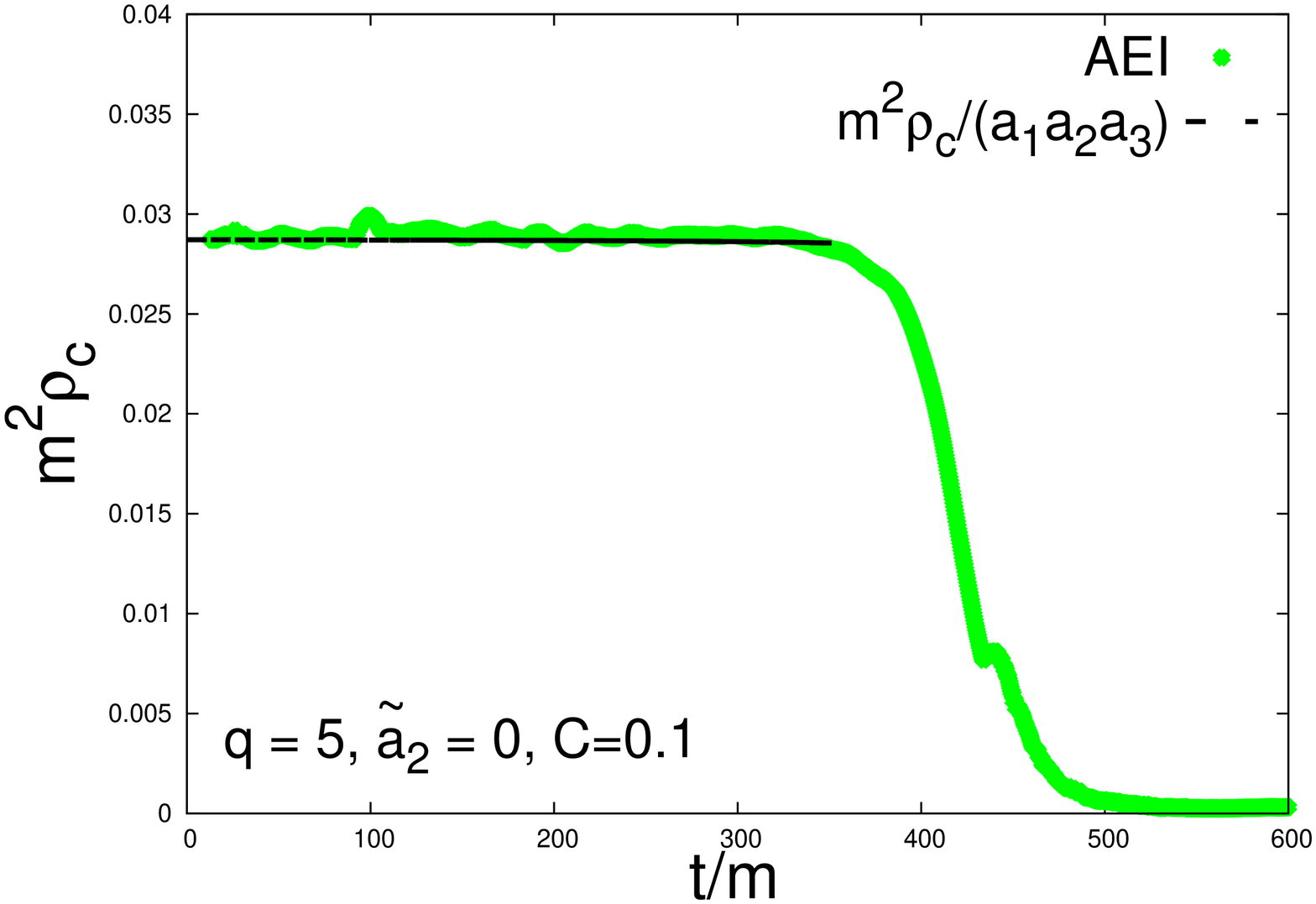,width=130pt,angle=270}%&
\epsfig{file=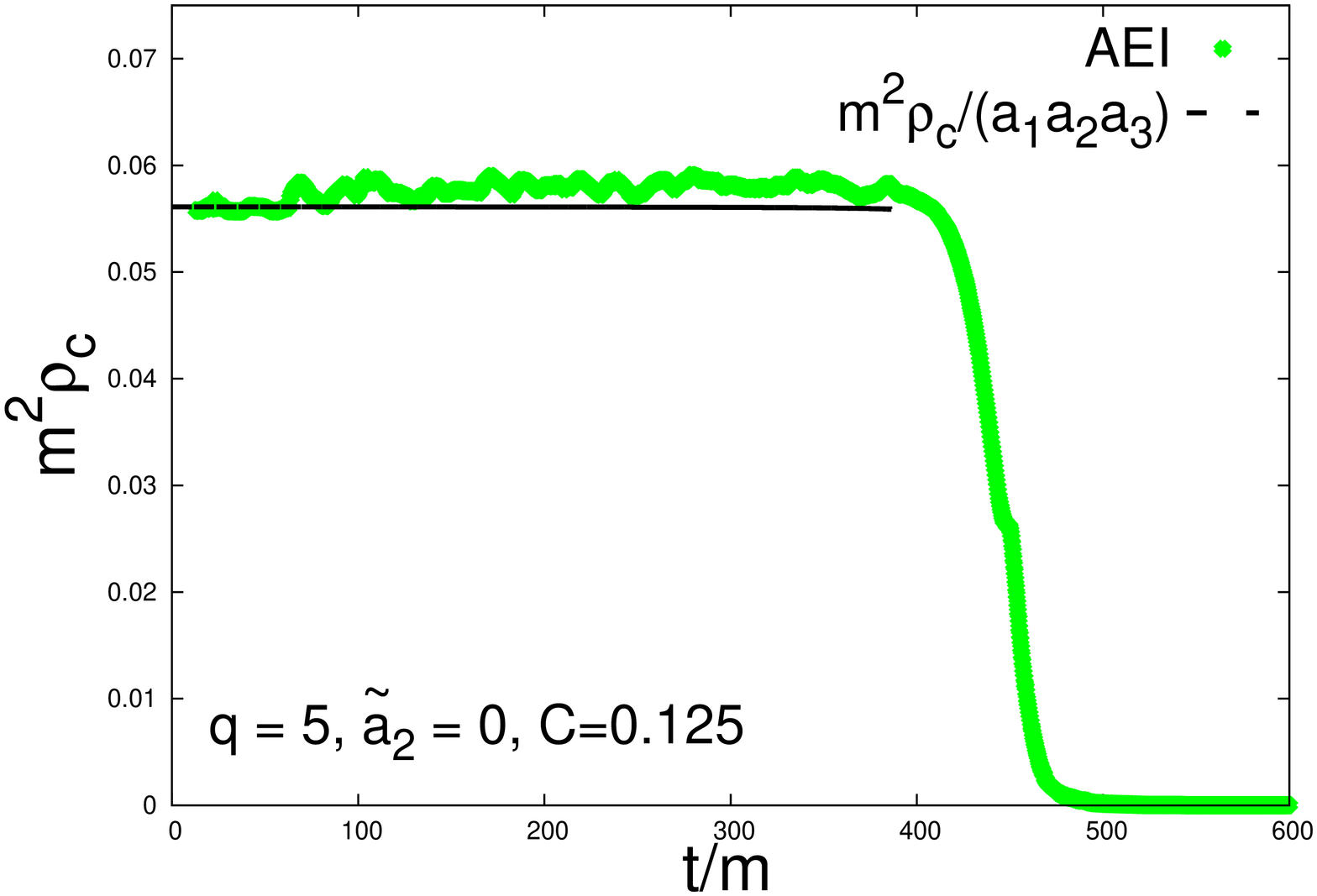,width=130pt,angle=270}\\
\epsfig{file=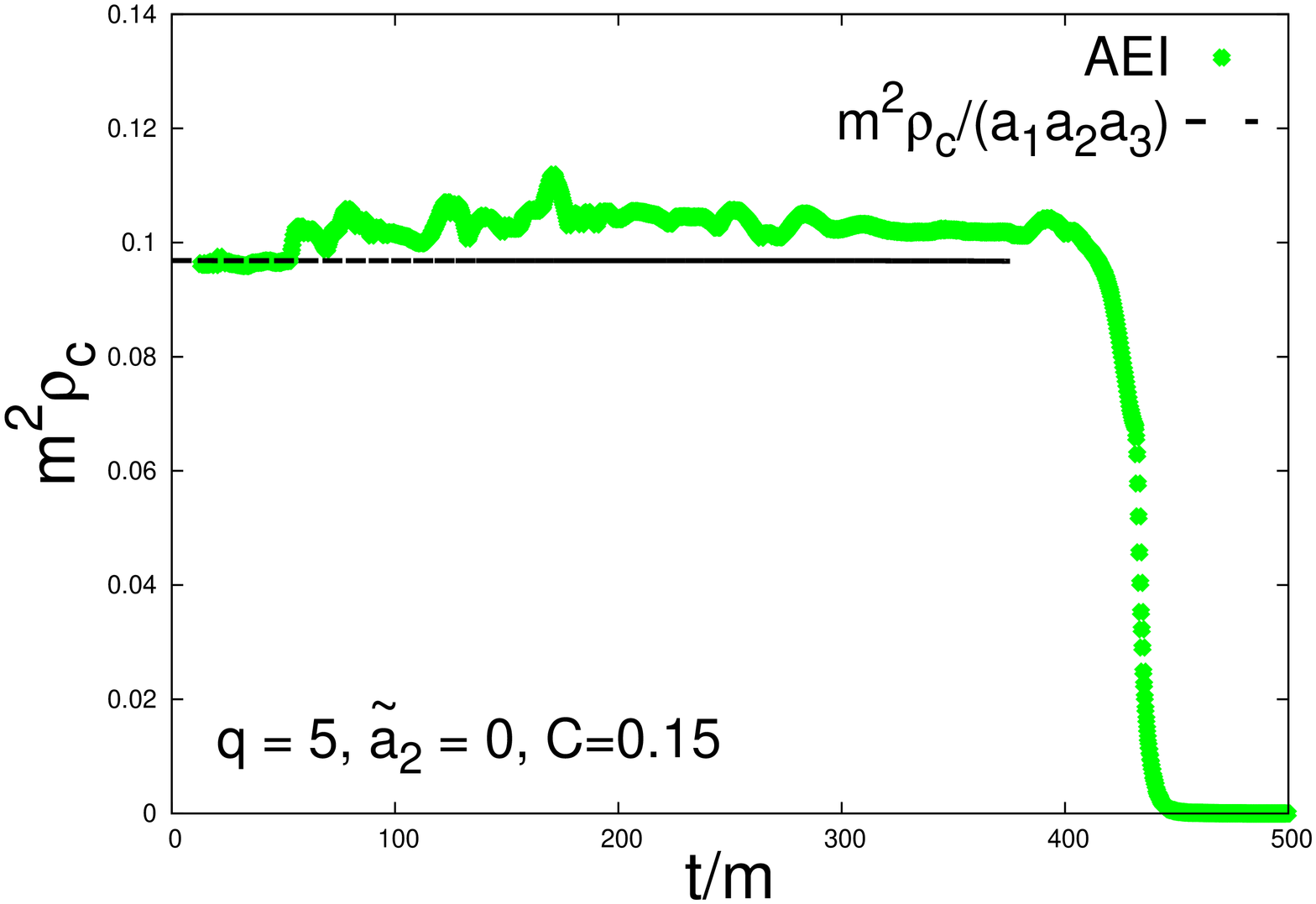,width=130pt,angle=270}%\\
\caption{(Color online) The NS central density (normalized to the squared total
  mass of the binary) is plotted, as a function of time (normalized to the
  binary total mass), for the AEI simulations (solid line) and for our
  simulations (dashed line).}
\label{fig_rho}
\end{figure*}

In order to assess the accuracy of our evaluation of the onset of mass shedding,
we consider the evolution of the NS central density, $\rho_c(t)$, which is a
gauge invariant quantity. We compare our profiles, with those evaluated by the
AEI group for the parameters indicated above.  The results are shown in
Fig.~\ref{fig_rho}, where the AEI profiles are plotted with a solid line, and
our data with a dashed line.  In the AEI curves, at some point the central
density sharply drops down.  This signals the transition to a new equilibrium
configuration and hence the possible transfer of mass from the star to the black
hole.  Determining accurately in the numerical simulations when this transfer
begins, is not trivial (tiny amounts of matter are lost from the stellar surface
already at large separations). However, it is reasonable to assume that the
transfer of mass takes place in the transition between the two different values
of the central density and therefore in a time interval of 0.75 ms for the C=0.1
model or of 0.25 ms for the C=0.15 model (clearly, the smaller the compactness
the slower the transfer process) \cite{privcomm}. Overall, therefore, we can
take the decrease of $\rho_c$ to roughly mark the stage in which the star fills
up the Roche lobe.  The dashed line for our $\rho_c(t)$ ends at the onset of
mass shedding. This occurs just before the steep decrease of the AEI curves, for
the models with compactness $C=0.1,0.125$.  For the case $C=0.15$, $r_{ICO}$ is
reached before mass-shedding sets in; therefore, the dashed line in the bottom
panel of Fig.~\ref{fig_rho} ends at an earlier time with respect to the sharp
drop of the solid line.  We also note that the values of the central density in
our simulations and in AEI's, agree quite well for the models with $C=0.1$ and
$C=0.125$. For the model with $C=0.15$, the AEI data show an increase of the
central density at earlier times, probably due to spurious numerical effects.

\begin{figure*}[t]
\epsfig{file=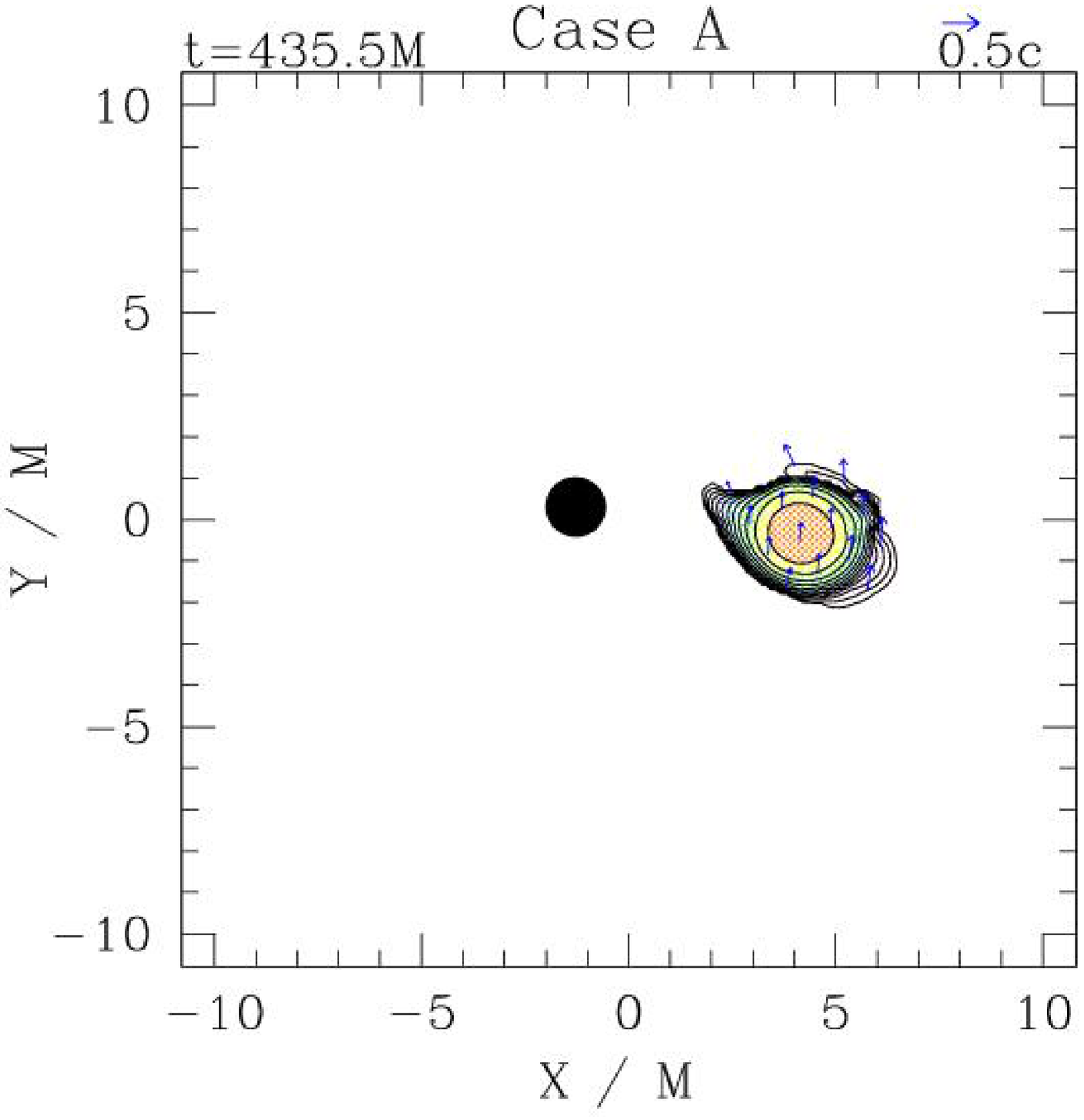,width=140pt}%&
\epsfig{file=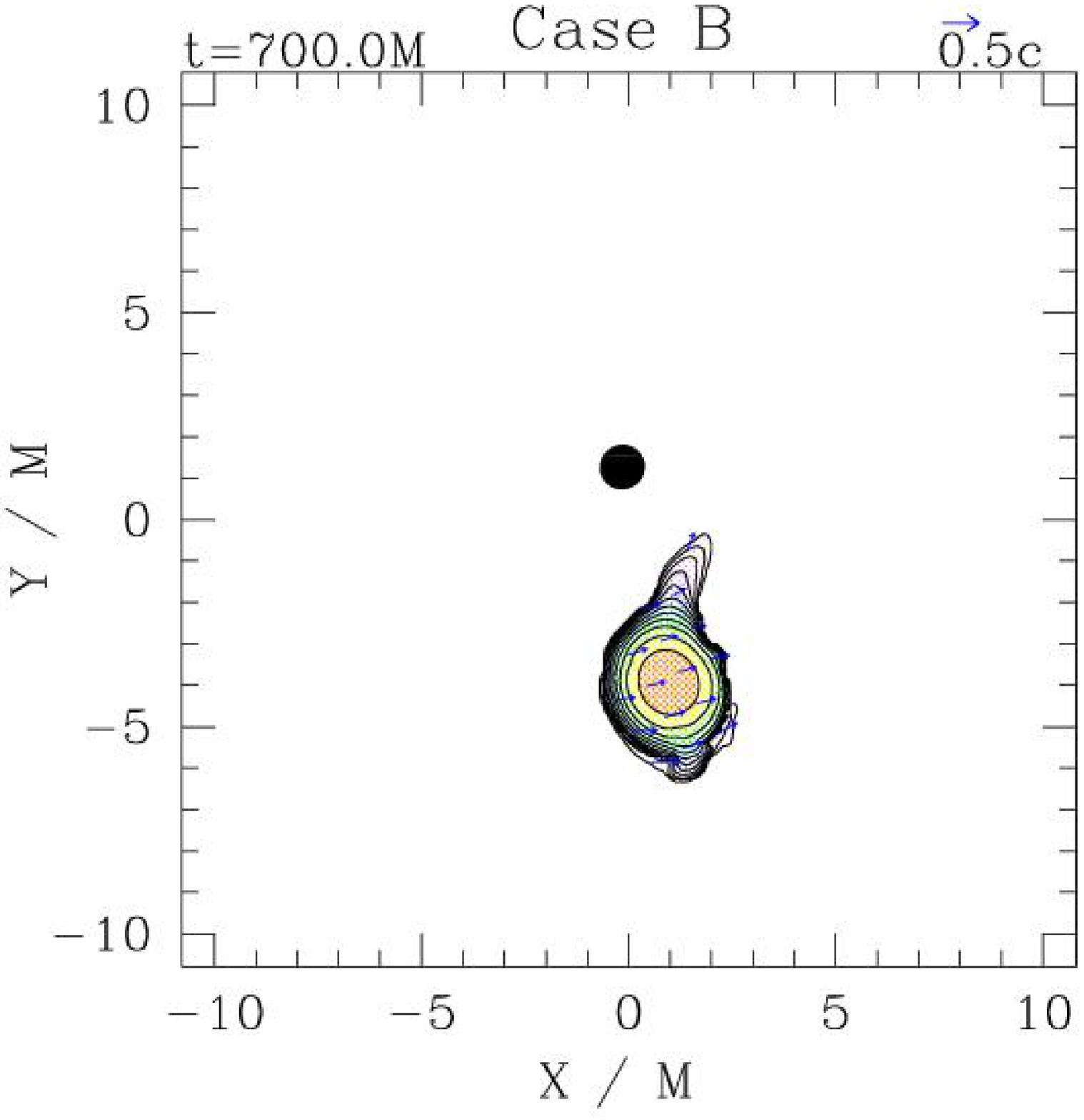,width=140pt}\\
\epsfig{file=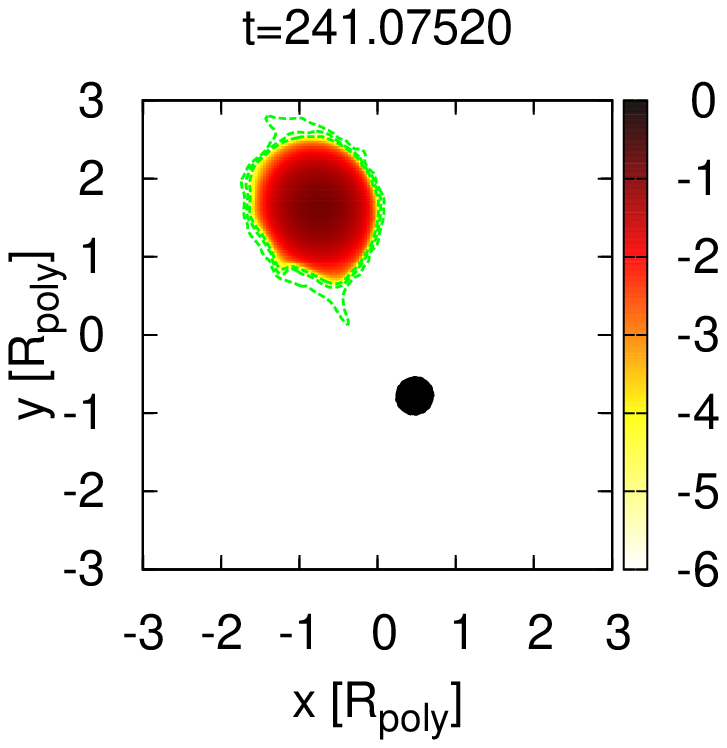,width=235pt}%&
\epsfig{file=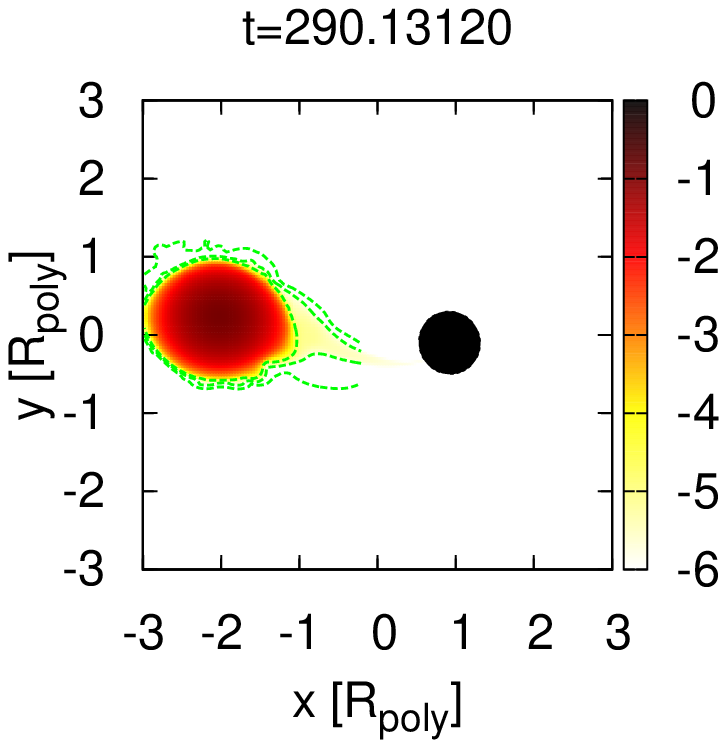,width=235pt}%\\
\caption{(Color online) Snapshots corresponding to $t=t_{shed}$ which we
evaluate integrating our equations. Upper panels:
  URB simulations with $C=0.145$, $\tilde a=0,0.75$, $q=3$.  Lower panels: KT
  simulations with $C=0.145$, $\tilde a=0$, $q=2,3$.}
\label{fig_snap}
\end{figure*}

As a further check, we have compared our estimate of the onset of mass shedding
with snapshots of the URB and KT simulations.  In Fig.~\ref{fig_snap}, we show
these snapshots at the time $t= t_{shed}$, which we evaluate for the different
models.  At that time, in the URB and KT simulations the NS starts showing a
cusp, indicating a mass flow.  We remark that this kind of comparison should be
considered as purely qualitative, since the stellar boundary in the snapshots
corresponds to a threshold value of the stellar density, the choice of which is
arbitrary.

%------------------------------------------------------------------------------
%-------------------------------------------------------------------------------
%-------------------------------------------------------------------------------
%-------------------------------------------------------------------------------
We would also like to remark that to compare the results of our approach
with those of fully relativistic simulations is not an easy task.  
In particular, a ``clean'' comparison of the stellar shape   
deserves further investigations in  close interaction with numerical
relativity groups, and will be considered in future works.

%%%%%%%%%%%%%%%%%%%%%%%%%%%%%%%%%%%%%%%%%%%%%%%%%%%%%%%%%%%%%%%%%%%%
\subsection{Love number}\label{seclove}
%%%%%%%%%%%%%%%%%%%%%%%%%%%%%%%%%%%%%%%%%%%%%%%%%%%%%%%%%%%%%%%%%%%%

A different kind of check can be performed by evaluating the Love number which,
as discussed in the Introduction, encodes the deformation properties of the
star.  When a weak tidal field induces a deformation on a spherical star, the
star (traceless) quadrupole moment is proportional to the tidal field
\cite{FH08,H08} (see also the generalizations discussed in \cite{BP09,DN09}):
\begin{equation} 
Q_{ij}=-\lambda c_{ij}=-\frac{2}{3}k_2R_{NS}^5c_{ij}\label{ad1}\,.
\end{equation} 
Eq.~(\ref{ad1}) (which is written in the principal frame) is based on the ``Love
number adiabatic approximation'' discussed in Section \ref{sec:affine} C; since
it assumes that the timescale of the orbital evolution (and then of the tidal
tensor changes) is much larger then that needed for the star to set into a
stationary configuration, this assumption may not be correct in the latest phase
of the inspiral; however, it is satisfied when the star and the companion are
sufficiently far apart.

In our model,  the NS quadrupole moment 
in the principal frame is
\begin{equation}
Q_{ij}=\frac{\hat{\cal M}}{R_{NS}^2}(a_ia_j-a^2\delta_{ij})\label{Ml}
\end{equation}
where $a^2\equiv(a_1^2+a_2^2+a_3^2)/3$, and $c_{ij}$ is given in
Eqns.~(\ref{cxx})-(\ref{cxy}).  In order to compare the Love number $k_2$
predicted by our approach with those determined by Hinderer \cite{H08}, which we
denote by $k_2^H$, we have evaluated $k_2$ using Eqns.~(\ref{ad1}), (\ref{Ml})
for the same NS models, assuming a polytropic equation of state with different
values of the adiabatic index $\Gamma$ and of the compactness $C$, by setting
the binary system at the orbital separation of $r_{12}\sim 180$ km.  As shown in
Table \ref{tablelove}, our results agree with those of \cite{H08} within a few
percent.

We remark that the Love number approach was generalized in \cite{BP09,DN09},
where other Love numbers were introduced; however, the leading tidal effect is
encoded in $k_2$.
%%%%%%%%%%%%%%%%%%%%%%%%%%%%%%%%%%%%%%%%%%%%%%%%%%%%%%%%%%%%%%%%%%%%%%%%%%%%%
\begin{table}[ht]
\begin{tabular}{ccccccc}
\hline
$C$ && $\Gamma$ && $k_{2}$ && $k_2^H$ \\  
\hline
0.10&&1.830&&0.0920&&0.0931\\
\hline 
0.15&&1.830&&0.0551&&0.0577\\
\hline 
0.20&&1.830&&0.0297&&0.0327\\
\hline 
0.10&\phantom{a}&2.000&\phantom{a}&0.1221&\phantom{a}&0.1220\\
\hline
0.15&&2.000&&0.0767&&0.0776\\
\hline
0.20&&2.000&&0.0444&&0.0459\\
\hline
0.10&&2.423&&0.1817&&0.1780\\
\hline
0.15&&2.423&&0.1198&&0.1170\\
\hline
0.20&&2.423&&0.0737&&0.0721\\
\hline
\end{tabular} 
\caption{The Love number $k_2$, evaluated  for different values of the
NS compactness $C$, and
  of the polytropic index $\Gamma$, is compared with the values obtained
in \cite{H08} for the same stellar models. \label{tablelove}}  
\end {table} 
%%%%%%%%%%%%%%%%%%%%%%%%%%%%%%%%%%%%%%%%%%%%%%%%%%%%%%%%%%%%%%%%%%%%%%%%%%%%%

%%%%%%%%%%%%%%%%%%%%%%%%%%%%%%%%%%%%%%%%%%%%%%%%%%%%%%%%%%%%%%%%%%%%%%%%%%%%%
\section{Concluding remarks}\label{conclusions}
%%%%%%%%%%%%%%%%%%%%%%%%%%%%%%%%%%%%%%%%%%%%%%%%%%%%%%%%%%%%%%%%%%%%%%%%%%%%%

In this article we have developed a post-Newtonian-affine (PNA) approach which
allows to model the tidal deformations of a neutron star in compact binary
coalescences.  To validate the model through a comparison with the results of
fully relativistic, numerical simulations, we have solved the dynamical
equations for BH-NS binary systems.  The tests we have made show a good
agreement with those results.

The PNA approach can be useful in many respects. It may complement numerical
relativity studies of binary coalescence because, due to its much lower
computational cost, it enables to study a large set of models, exploring a wide
range of parameters.  Furthermore, like all semi-analytic approaches, it would
be helpful to dig the physical features of the process out of the numerical
artifacts which may affect the fully relativistic simulations.

Since the PNA approach does not assume the ``Love number adiabatic
approximation'', it would allow to test the validity domain of this
assumption. Indeed, using the PNA framework, it would be possible to determine
under which conditions the NS deformation is characterized by a set of constant
coefficients, and to find their behaviour, if they change during the inspiral.

Finally, we would like to remind that the production of initial data for fully
relativistic simulations is a very delicate task.  Typically, initial data are
plagued by spurious effects like, for instance, a non-physical eccentricity
(compact binaries are known to circularize well ahead the latest stages of the
inspiral); it is difficult to produce truly general initial data (for instance,
with non-aligned spins).  The PNA approach could be used to produce initial data
for fully relativistic simulations, complementing existing initial data solvers.

These aspects, and the extension of the PNA approach to the study of NS-NS
coalescences, will be the matter of future works.

%%%%%%%%%%%%%%%%%%%%%%%%%%%%%%%%%%%%%%%%%%%%%%%%%%%%%%%%%%%%%%%%%%%%%%%%%%%%%
\section*{Acknowledgments}
%%%%%%%%%%%%%%%%%%%%%%%%%%%%%%%%%%%%%%%%%%%%%%%%%%%%%%%%%%%%%%%%%%%%%%%%%%%%%% 
We thank F. Pannarale, L. Rezzolla, U. Sperhake, Z. Etienne and K.
Kyutoku  for useful discussions. We
thank the Albert Einstein Institute Numerical Relativity Group \cite{privcomm},
the Illinois Numerical Relativity Group \cite{E09}, and the Numerical Relativity
Groups of the Kyoto and Tokyo Universities \cite{S09}, for kindly sharing their
data with us. This work was partially supported by CompStar, a research
networking program of the European Science Foundation.  L.G. has been partially
supported by Grant No. PTDC/FIS/098025/2008.

\appendix
%%%%%%%%%%%%%%%%%%%%%%%%%%%%%%%%%%%%%%%%%%%%%%%%%%%%%%%%%%%%%%%%%%%%%%%%%%%%%
\section{Post-Newtonian expressions for the orbital motion}\label{appa}
%%%%%%%%%%%%%%%%%%%%%%%%%%%%%%%%%%%%%%%%%%%%%%%%%%%%%%%%%%%%%%%%%%%%%%%%%%%%%% 
In this Appendix we write explicitly the post-Newtonian coefficients $a_{k}$ 
of the Taylor T4 approximant eq.(\ref{def:TaylorT4a}), for spinning bodies in
quasi-circular orbits, with spins aligned with the direction of the Newtonian
orbital angular momentum vector \cite{SOA10}:
\begin{widetext}
{\small
\begin{eqnarray} 
 a_0 &=& 1 , \quad a_1 = 0 , \quad a_2 = -\frac{743}{336}-\frac{11 \nu }{4} , 
\quad a_3 = 4 \pi - \frac{113}{12}  \chi + \frac{19 \nu}{6}  (\tilde{a}_{1} + \tilde{a}_{2} ) , \nonumber \\  
a_4 &=& \frac{34103}{18144} + 5 \chi^2  + \nu \left( \frac{13661}{2016} - 
\frac{\tilde{a}_{1} \tilde{a}_{2}}{8}\right) + \frac{59 \nu ^2}{18} , \label{eq:T4Coeffs} \nonumber \\ 
a_5 &= &- \pi \left(\frac{4159 }{672} + \frac{189}{8} \nu\right) -\chi 
\left( \frac{31571}{1008} - \frac{1165}{24} \nu\right)  + (\tilde{a}_{1} + 
\tilde{a}_{2}) \left( \frac{21863}{1008} \nu - \frac{79}{6}\nu^2\right) - 
\frac{3}{4} \chi^3 + \frac{9 \nu}{4} \chi  \, \tilde{a}_{1} \tilde{a}_{2}\ ,\nonumber    \\ 
a_6 &=& \frac {16447322263}{139708800} - \frac {1712}{105}\, \gamma_E 
+\frac{16 \pi^{2}}{3}-\frac {856}{105} \ln  \left(16 x \right)  + \nu 
\left (\frac {451 {\pi}^{2}}{48} - \frac {56198689}{217728} \right )+{\frac {541}{896}}\,{\nu}^{2}
-{ \frac {5605}{2592}}\,{\nu}^{3}\nonumber \\
&&-\frac{80 \pi}{3} \chi + \left( \frac{20 \pi}{3} - \frac{1135}{36} \chi 
\right)\nu (\tilde{a}_{1} + \tilde{a}_{2})+ \left( \frac{64153}{1008} -
  \frac{457 }{36} \nu  \right) 
\chi^2 - \left( \frac{787 }{144} \nu  - \frac{3037 }{144} \nu ^2 \right) \tilde{a}_{1} \tilde{a}_{2}\ , 
   \nonumber \\
 a_7 &=&- \pi \left( \frac {4415}{4032} -\frac {358675}{6048} \nu
  -\frac {91495}{1512} \nu^{2} \right) 
- \chi \left( \frac{2529407}{27216} - \frac{845827 }{6048} \nu +
\frac{41551}{864} \nu ^2 \right)+ 12
\pi  \chi^2 - \chi^3 \left(\frac{1505}{24} + \frac{\nu}{8} \right)\nonumber \\
&& \quad + (\tilde{a}_{1} + \tilde{a}_{2}) \left( \frac{1580239 }{54432} \nu
-\frac{451597}{6048}  \nu ^2 + \frac{2045}{432}  \nu ^3 
+\frac{107 \nu  }{6} \chi ^2 -\frac{5 \nu ^2}{24} \tilde{a}_{1} \tilde{a}_{2}
\right)  + \chi  \tilde{a}_{1} \tilde{a}_{2} \left( \frac{101  }{24} \nu + \frac{3}{8} \nu^2 \right)
 \nonumber 
\end{eqnarray}}
\end{widetext}
Here $\gamma_{E}$ is Euler's constant and $\chi =\frac{m_1}{m}\tilde{a}_{1}+\frac{m_2}{m}\tilde{a}_{2}$, 
with $\tilde{a}_{1,2}$ dimensionless spin parameters defined in section \ref{orbital}.
%%%%%%%%%%%%%%%%%%%%%%%%%%%%%%%%%%%%%%%%%%%%%%%%%%%%%%%%%%%%%%%%%%%%%%%%%%%%%
\section{The tidal tensor}\label{tidalcij}
%%%%%%%%%%%%%%%%%%%%%%%%%%%%%%%%%%%%%%%%%%%%%%%%%%%%%%%%%%%%%%%%%%%%%%%%%%%%%% 
We show the non-vanishing components of the tidal tensor $C^{i}_{j}$ up to the
$1/c^{3}$ order, as functions of the PN potentials:
\begin{widetext}
{\small
\begin{eqnarray}
  C_{xx}=&-&\pxx\vo+\frac{1}{c^{2}}\Bigg\{-4\pxt\vx+4v^{y}\pxx\vy-
4 v^{y}\pxy\vx-(\py\vo)^{2}-\pxx\voo-\bigg[\ptt +(v^{y})^{2}(\pyy+2\pxx)\ +\nonumber\\ 
  &+&2v^{y}\pyt-v^{x}v^{y}\pxy\bigg]\vo+2\pxx\vo\vo+2(\px\vo)^{2}
\Bigg\}-\frac{4}{c^{3}}\left\{(\pxt+v^{y}\pxy)\vxx-v^{y}\pxx\vyy+\frac{1}{4}\partial^{2}_{xx}V^{(3)}\right\}\ ,\\
  C_{yy}=&-&\pyy\vo+\frac{1}{c^{2}}\Bigg\{-4\pyt\vy-(\px\vo)^{2}+4v^{x}
\pyy\vx -4v^{x}\pxy\vy-\pyy\voo-\bigg[\ptt+(v^{x})^{2}(\pxx+2\pyy)\ +\nonumber\\
  &+&2v^{x}\pxt-v^{x}v^{y}\pxy\bigg]\vo+2\pyy\vo\vo+2(\py\vo)^{2}
\Bigg\}-\frac{4}{c^{3}}\left\{(\pyt+v^{x}\pxy)\vyy-v^{x}\pyy\vxx+\frac{1}{4}\partial^{2}_{yy}V^{(3)}\right\}\ , \\ 
  C_{zz}=&-&\pzz\vo+\frac{1}{c^{2}}\Bigg\{4v^{y}\pzz\vy+4v^{x}\pzz\vx-
\pzz\voo-(\px\vo)^{2}-(\py\vo)^{2}-\bigg[\ptt+2\left[(v^{x})^{2}+(v^{y})^{2}\right]\pzz\ 
+\nonumber\\
  &+&2v^{x}\pxt+2v^{y}\pyt+(v^{x})^{2}\pxx+(v^{y})^2\pyy+2v^{x}v^{y}\pxy\bigg]\vo+2\pzz
\vo\vo\Bigg\}\ +\nonumber\\
&+&\frac{4}{c^{3}}\left\{v^{y}\pzz\vyy+v^{x}\pzz\vxx-\frac{1}{4}\partial^{2}_{zz}V^{(3)}\right\}\ ,\\
  C_{xy}=&-&\pxy\vo+\frac{1}{c^{2}}\Bigg\{\bigg[v^{y}\pxt+v^{x}\pyt+
\frac{3}{2}v^{x}v^{y}(\pxx+\pyy)-\frac{(v^{x})^2+
(v^{y})^2}{2}\pxy\bigg]\vo+3\px\vo\py\vo+\pxy\voo\ + \nonumber\\
  &+&2\pxy\vo\vo+2(v^{x}\pxy-v^{y}\pyy-\pyt)\vx+2(v^{y}\pxy
-v^{x}\pxx-\pxt)\vy\Bigg\}\ +\nonumber\\
  &+&\frac{2}{c^{3}}\left\{(v^{x}\pxy-v^{y}\pyy-\pyt)\vxx+(v^{y}\pxy-v^{x}\pxx-\pxt)\vyy-\frac{1}{2}\partial^{2}_{xy}V^{(3)}
\right\}\ .
\end{eqnarray}}
\end{widetext} 

%%%%%%%%%%%%%%%%%%%%%%%%%%%%%%%%%%%%%%%%%%%%%%%%%%%%%%%%%


\begin{thebibliography}{99}
\bibitem{virgoligo} {\tt http://www.ego-gw.it; http://www.ligo.caltech.edu}.
\bibitem{NPP09} R.~Narayan, B.~Paczynski, and T.~Piran, Astrophys.\  J.\
  Lett. {\bf 395}, L83 (1992).
\bibitem{B06} L.~Blanchet, Living Rev. Relativity, {\bf 9}, 4 (2006).
\bibitem{EOB} A. Buonanno, T. Damour, Phys.Rev. D59 (1999) 084006;
T.~Damour, A.~Nagar, Phys.\ Rev. D {\bf 79}, 081503 (2009).
\bibitem{MW04} T.~Mora, C.M.~Will, Phys. Rev. {\bf D69}, 104021 (2004).
\bibitem{FH08} E.E.~Flanagan, T.~Hinderer, Phys. Rev. {\bf D77}, 021502 (2008).
\bibitem{H08} T.~Hinderer, Astrophys.\ J.\ {\bf 677}, 1216 (2008); {\it ibid.},
  {\bf 697}, 964 (2009).
\bibitem{VFH11} J.~Vines, E.E.~Flanagan, T.~Hinderer, Phys.\ Rev. {\bf D83}, 084051 (2011).
\bibitem{EOBtidal} T. Damour, A. Nagar, Phys.Rev. D81 (2010) 084016.
\bibitem{F09} J.~A.~Faber, Class.\ Quant.\ Grav.\  {\bf 26}, 114004
  (2009).
\bibitem{TBFS07} K.~Taniguchi, T.~W.~Baumgarte, J.~A.~Faber, S.~L.~Shapiro,
Phys.\ Rev.\  D {\bf 75}, 084005 (2007).
\bibitem{S09q} K.~Kyutoku, M.~Shibata, and K.~Taniguchi,
Phys. Rev. {\bf D79}, 124018 (2009).
\bibitem{FGP09} V.~Ferrari, L.~Gualtieri, F.~Pannarale, Class.\ Quant.\ Grav.\
  {\bf 26}, 125004 (2009).
\bibitem{FGP10} V.~Ferrari, L.~Gualtieri, F.~Pannarale, Phys.\ Rev.\ {\bf D81},
  064026 (2010).
\bibitem{F06a} J.~A.~Faber, T.~W.~Baumgarte, S.~L.~Shapiro, K~Taniguchi,
F.~A.~Rasio, Phys.\ Rev.\  D {\bf 73}, 024012 (2006).
\bibitem{F06b} J.~A.~Faber, T.~W.~Baumgarte, S.~L.~Shapiro, K~Taniguchi,
Astrophys.\ J. {\bf 641}, L93 (2006).
\bibitem{S07} M.~Shibata, K.~Uryu,Class.\ Quant.\ Grav.\  {\bf 24}, S125 (2007).
\bibitem{S09} M.~Shibata, K.~Kyutoku, T.~Yamamoto, and K.~Taniguchi,
 Phys. Rev. {\bf D79}, 044030 (2009).
\bibitem{S10} K.~Kyutoku, M.~Shibata, and K.~Taniguchi,
Phys. Rev. {\bf D82}, 044049 (2010); {\bf 84} 049902 (2011).
\bibitem{E09} Z.~B.~Etienne, Y.~T.~Liu, S.~L.~Shapiro,T.~ W.~Baumgarte, 
Phys.\ Rev.\  D {\bf 79}, 044024 (2009).
\bibitem{S11} K.~Kyutoku, H.~Okawa, M.~Shibata, and K.~Taniguchi,
Phys. Rev. {\bf D84}, 064018 (2011).
\bibitem{ST11} M.~Shibata, K.~Taniguchi, Living Rev. Relativity, {\bf 14}, 6 (2011).
\bibitem{CL85} B.~Carter, J.P.~Luminet, Mon.\ Not.\ R.\ Astron.\ Soc. {\bf 212}, 23 (1985).
\bibitem{LM85}  J.P.~Luminet, J.A.~Marck, Mon.\ Not.\ R.\ Astron.\ Soc. {\bf 212}, 57 (1985).
\bibitem{WL00} P.~Wiggins,  D.~Lai,  Astrophys.\ J. {\bf 532}, 530 (2000).
\bibitem{CFS06} C.~Casalvieri, V.~Ferrari, A.~Stavridis, Mon.\ Not.\ R.\
  Astron.\ Soc. {\bf 365}, 929 (2006).
\bibitem{DSX92} T.~Damour, M.~Soffel, C.~Xu, Phys.\ Rev. {\bf D45}, 1017 (1992). 
\bibitem{RF05} E.~Racine, E.E.~Flanagan, Phys.\ Rev. {\bf D71}, 044010 (2005).
\bibitem{VF09} J.~Vines, E.E.~Flanagan, arXiv:1009.4919
\bibitem{PROR11} F.~Pannarale, L.~Rezzolla, F.~Ohme, J.~Read,  Phys.\ Rev. {\bf  D84}, 104017 (2011).
\bibitem{PTR11} F.~Pannarale, A.~Tonita, L.~Rezzolla, Astrophys. \ J. {\bf 727}, 95 (2011).
\bibitem{OK96} W.~Ogawaguchi, Y.~Kojima, Prog.\ Theor.\ Phys. {\bf 96}, 901 (1996).
\bibitem{HLLR10} T.~Hinderer, B.D.~Lackey, R.N.~Lang, J.S.~Read, Phys. Rev. {\bf D81}, 123016 (2010).
\bibitem{DN09} T.~Damour, A.~Nagar, Phys.\ Rev. {\bf D80}, 084035 (2009).
\bibitem{BP09} T.~Binnington, E.~Poisson, Phys.\ Rev. {\bf D80}, 084018 (2009).
\bibitem{DN10} T.~Damour, A.~Nagar, Phys.\ Rev. {\bf D81}, 084016 (2010).
\bibitem{privcomm} L.~Rezzolla, F.~Pannarale, private communication (2011).
\bibitem{EFE} S.~Chandrasekhar {\it Ellipsoidal Figures of Equilibrium (The
    Silliman Foundation Lectures)} (Yale University Press, New Haven) (1969).
\bibitem{Chan65} S.~Chandrasekhar, Astrophys. J {\bf 142}, 1488 (1965).
\bibitem{BFP98} L.~Blanchet, G.~Faye, and B.~Ponsot, Phys. Rev. {\bf D58}, 124002 (1998).
\bibitem{FBA06} G.~Faye, L.~Blanchet, and A.~Buonanno, Phys. Rev. {\bf D74}, 104033 (2006).
\bibitem{BAF11}  L. Blanchet, A. Buonanno, and G. Faye, Phys. Rev. {\bf D84}, 064041 (2011).
\bibitem{P64} P.C.~Peters, Phys. Rev. {\bf 136}, B1224 (1964).
\bibitem{DIST01} T. Damour, B.R. Iyer, B.S. Sathyaprakash, Phys.Rev. D63 044023 (2001).
\bibitem{SOA10}L.~Santamaria et al., Phys. Rev. {\bf D82}, 064016 (2010).
\bibitem{BBK07} M.~Boyle, D.A.~Brown, L.E.~Kidder, A.H.~Mroue, H.P.~Pfeiffer,
 M.A.~Scheel, G.B.~Cook, S.A.~Teukolsky, Phys. Rev. {\bf D76}, 124038 (2007).
\bibitem{Fav11} Marc Favata, Phys. Rev. D {\bf 83},024027 (1011).
\bibitem{M75} B.~Mashhoon, Astrophys.\ J. {\bf 197}, 705 (1975).
\bibitem{M83} J.-A.~Marck, Proc.\ Roy.\ Soc.\ Lond. {\bf Q385}, 431, (1983).
\bibitem{P56} F.A.E.~Pirani, Acta phys.pol. {\bf 15}, 389 (1956).
\bibitem{F88} T.~Fukushima, Celest.\ Mech. {\bf 75}, 4461 (1988).
\bibitem{BI03} L.~Blanchet, B.R.~Iyer, Class.\ Quant.\ Grav. {\bf 20}, 755 (2003).
\bibitem{FDB00} V.~Ferrari, M.~D'Andrea, E.~Berti, Int.\ J.\ Mod.\ Phys. {\bf
    D9}, 495 (2000).
\bibitem{L94} D.~Lai, Mon.\ Not.\ R.\ Astron.\ Soc. {\bf 270}, 611 (1994).
\bibitem{HHGSB08} M.~Hassan, S.~Husa, J.A.~Gonzalez, U.~Sperhake, B.~Br\"ugmann, 
Phys. Rev. {\bf D77}, 044020 (2008).
\bibitem{CLNZ09} M.~Campanelli, C.O.~Lousto, H.~Nakano, Y. Zlochower, 
Phys. Rev. {\bf D79}, 084010 (2009).
\bibitem{LKSB11} B.D. Lackey, K. Kyutoku, M. Shibata, P.B. Brady, J.L. Friedman, 
arXiv:1109.3402.
\bibitem{BDGNR11} L. Baiotti, T. Damour, B. Giacomazzo, A. Nagar, L. Rezzolla, 
Phys.Rev. {\bf D84} (2011) 024017, Phys.\ Rev.\ Lett.\ {\bf 105} 261101 (2010).
\end{thebibliography}
\end{document}